\documentclass[]{aa}
\usepackage{graphicx}
\usepackage{txfonts}
\usepackage{natbib}
\bibpunct{(}{)}{;}{a}{}{,}   
\usepackage{xspace}

\newcommand{\rxj}{RX J1713.7-3946\xspace}
\newcommand{\g}{$\gamma$\xspace}
\newcommand{\x}{\textit{XMM-Newton}\xspace}

\newcommand{\kms}{$\rm km\,s^{-1}$}

\begin{document}

   \title{A joint spectro-imaging analysis of the \x and HESS observations of the supernova remnant RX J1713.7-3946}

 \authorrunning{Acero et al.}
 \titlerunning{An X- and Gamma-ray comparison of RX J1713.7-3946}

   \author{F. Acero\inst{1}, J. Ballet\inst{1}, A. Decourchelle\inst{1}, M. Lemoine-Goumard\inst{2,}\inst{3}, M. Ortega\inst{4}, E. Giacani\inst{4}, G. Dubner\inst{4}, G. Cassam-Chena\"i\inst{5}. }


   \institute{1: Laboratoire AIM, CEA/DSM-CNRS-Universit\'e Paris Diderot, IRFU/SAp, CEA-Saclay, 91191 Gif sur Yvette, France \\
\email{fabio.acero@cea.fr}\\
2: CNRS/IN2P3, Centre d'Etudes nucl\'eaires de Bordeaux Gradignan, UMR 5797, Gradignan 33175, France \\
3: Universit\'e Bordeaux I,  Centre d'Etudes nucl\'eaires de Bordeaux Gradignan, UMR 5797, Gradignan 33175, France \\
4: Instituto de Astronom\'ia y F\'isica del Espacio (IAFE), CC 67, Suc. 28, 1428 Buenos Aires, Argentina\\
5: INAF-Osservatorio Astrofisico di Arcetri, Largo E. Fermi, 5, 50125 Firenze, Italy   }

\date{Received 19 december 2008 / Accepted 31 May 2009}

  \abstract
   {The supernova remnant (SNR) \rxj (also known as G347.3-0.5) is part of the class of remnants dominated by synchrotron
   emission in X-rays. It is also one of the few shell-type SNRs observed at TeV energies allowing to
investigate particle acceleration at SNRs shock.
    }
   {Our goal is to compare spatial and spectral properties of the remnant in X- and \g-rays 
to understand the nature of the TeV emission.
This requires to study the remnant at the same spatial scale at both energies. 
To complement the non-thermal spectrum of the remnant, 
we attempt to provide a reliable estimate for the radio flux 
density.}
   {In radio, we revisited ATCA data and used HI and mid-infrared observations to
disentangle the thermal  from the non-thermal emission.
   In X-rays, we produced a new mosaic of the remnant and degraded the spatial resolution
   of the X-ray data to the resolution of the HESS instrument to perform spatially resolved 
spectroscopy at the same spatial scale in X- and \g-rays. Radial profiles were obtained to investigate the extension
   of the emission at both energies.}
   {We found that part of the radio emission within the SNR contours is thermal in nature.
	 Taking this into account, we provide  new lower and  upper limits for the integrated synchrotron flux of
	 the remnant at 1.4 GHz of 22 Jy and 26 Jy respectively.
In X-rays, we obtained the first full coverage of \rxj with \x.
	The spatial variation of the photon index seen at small scale in X-rays is
	smeared out at HESS resolution. A non-linear  correlation between the X- and
 	\g-ray fluxes of the type $F_{\mathrm{X}} \propto F_{\mathrm{\gamma}}^{2.41}$ is found.
  If the flux variation are mainly due to density  variation around the remnant then a leptonic model can more easily 
reproduce the observed X/\g-ray correlation.
	In some angular sectors, radial profiles indicate that
  	the bulk of the X-ray emission  comes more from the inside of the remnant than in \g-rays. }
   {}


   \maketitle

\section{Introduction}

Supernova remnants (SNRs) have long been believed to be accelerators of cosmic rays at least up to the \textit{knee}
($\sim10^{15}$ eV). Evidence that electrons are indeed accelerated in SNRs is found both in radio 
and X-rays through synchrotron emission.
The detection of TeV emission from a SNR is an evidence of proton acceleration
 if the \g-rays come from the interaction of
accelerated protons with the ambient matter (hadronic model). However the  \g-rays can also be produced via Inverse
Compton scattering (IC) of accelerated TeV electrons off ambient photons (leptonic model) that could be either infrared
surrounding emission or cosmic microwave background.
The SNRs interacting with dense molecular clouds are good candidates to detect \g-ray emission
from the hadronic mechanism as the high density of the cloud provides a large amount of targets
for the  accelerated protons.
The SNR \rxj (also known as G347.3-0.5)  is one of those candidates as it is interacting
 with a dense molecular cloud in the northwest (NW)
and in the southwest (SW) of the remnant \citep{fm03,cdg04,mt05}.

\rxj was first discovered in X-rays with the ROSAT all-sky survey in 1996 \citep{pa96}.  The remnant
is close to the galactic plane and its distance is controversial. Using ASCA observations and the measurement of
the column density toward the source,
it was first proposed to be at a distance of 1 kpc \citep{kk97}. Then \citet{sl99} derived a distance of 6 kpc based on a possible
association of the remnant with molecular clouds and the HII region G347.611+0.204.
The comparison of the X-ray and HI absorbing column densities \citep{cdg04} as well as CO observations
 of a  molecular cloud interacting with the remnant \citep{fm03} both suggest a closer distance of 
1 kpc (value
 adopted in this paper).
In this case, the remnant is about 20 pc in diameter (70 arcmin on the sky) and could be associated with
the supernova that exploded in AD 393 in the tail of the constellation Scorpius \citep{wq97}.
The remnant would then be about 1600 years old. \\
In the radio band \citet{sl99} showed for the first time an
image of RX J1713.7-3946, based on MOST observations
at 843 MHz.
Later on, \citet{es01} and \citet{ls04} (with a recalibrated and  improved  image) reported ATCA radio
observations of this SNR at 1.4 GHz (left-hand
panel of Fig. \ref{fig:radioir}).  The SNR appears as a faint nebula ($\sim$
1$^{\circ}$ in diameter) with many short, curved features, the brightest 
of which are the two bright arcs visible on the northwest corner.
These arcs coincide with the edges of the brightest X-ray emission. 
Interestingly, those arcs are observed to be in the vicinity of the HII
region G347.611+0.204 which is located at 6.6 kpc \citep{ru03,cc04}.
An almost complete ring of weak emission, 
about 30 arcmin in size, can be also detected near the center of this 
extended SNR.
An accurate estimate of the integrated radio flux density of
\rxj  is a pending problem. 
This is due to the intrinsic faintness 
of this SNR, the possible mixing of the SNR synchrotron emission with the thermal 
emission of the nearby HII region  as well as the limited quality of the available data.

In X-rays the emission is dominated by a non-thermal continuum and no emission lines
have been observed so far \citep{kk97,sl99,pa03,cdg04}. This non detection can set an upper limit to
the ionization age. Assuming $t_{SNR}=1600$ yrs an upper limit on the density of the ambient medium of 0.02 cm$^{-3}$ is then derived \citep{cdg04}.
 The study of the remnant with \x at  small spatial scale carried out by \citet{cdg04} and \citet{hu05}  clearly
showed a spatial variation of the photon index.
It varies from 1.9 to 2.6 with a mean statistical error of 4\% \citep{cdg04}.
 A  correlation between the X-ray flux and the photon index is also observed. Those results were obtained using
an adaptive grid to have approximately the same number of counts in each pixel grid region
where a spectrum is then extracted. On the bright regions (the  north-west rim),
 the size of the grid pixel is typically 0.05$\degr$. \\
 In \g-rays, \rxj has been detected
 by the CANGAROO collaboration \citep{mu00,en02}. It was then definitely  established by the
 HESS telescopes which  provided the first spatially resolved \g-ray image of the remnant \citep{ah04}.
 The overall morphology is very similar to that in X-rays (same western shell)
and the brightest spots (located to the north-west) are coincident.
As the remnant is extended and is the brightest shell-type SNR seen in \g-rays, its emission can be studied in detail.
With the good spatial resolution of the HESS telescopes ($\sim$ 0.1 $\degr$)
 it has been possible to carry out a spatially resolved spectral study  at large scale \citep[][ hereafter AH06]{ah06}.
No evidence for spatial variations of the photon index from region to region (mean value of 2.09 and a standard deviation of 0.07) was found down 
to the precision of the measurements (mean 1$\sigma$ statistical error of $\pm$ 0.08).
Moreover there does not seem to be any correlation between the \g-ray flux and the photon index 
(Fig. 14 of AH06). \\
 Those \g-ray results are very different from what has been found at small
scale in X-rays.
However it is important to note that the  X- and  \g-rays results 
 were not obtained looking at the same spatial scale
(the spectra were not extracted using the same size of extraction regions).
Moreover the Point Spread Function of both instruments is very different.
Those differences hamper a comparison of the results at both energies.
The major issue here is  whether the spectral parameters are really different in X- and \g-rays or if this difference
 is introduced by the fact that both studies are looking at different spatial scales. \\
To investigate this question we carried out a detailed comparison of the remnant  in X- and \g-rays using the
same extraction regions and taking into account the  different spatial resolution of the two instruments.
We compared both the  spectral properties of the remnant and the morphology
 using radial profiles in X- and \g-rays. \\
Concerning the radio flux estimate, we investigated which part of the radio emission
 is synchrotron emission in origin (related to the remnant) or thermal emission 
(possibly related to the nearby HII region). 
We then calculated the integrated flux density  re-analysing Lazendic et al. (2004)'s database. \\
In Sect. \ref{sect:data}, we present the different sets of data used in this study including the new
\x observations that complete the mosaic of the remnant. Section \ref{sect:xrproc} presents
 the  methods used for the processing of the X-ray data.  In Sect. \ref{sect:results}, 
the estimate of the integrated radio flux of the remnant and the results of the X- and \g-ray spectral and morphological comparison are shown. In Sect. \ref{sect:discu}, we discuss this comparison in the 
framework of an hadronic and leptonic model. The implications of the new radio flux for the 
multi-wavelength emission models of the remnant are also discussed.

\section{Data}
\label{sect:data}

\subsection{Radio data}
\label{sect:radiodata}
In this work we re-analyzed ATCA observations carried out at 1.4 GHz
\citep{ls04} following different paths to estimate the radio
flux density of RX J1713.7-3946.  The details to carry out an accurate calculation 
are discussed in Sect. \ref{sect:radioflux}.

\subsection{Infrared data}
\label{sect:irdata}

Infrared observations in and around the remnant can help 
 disentangling the thermal from the non thermal radio emission which is crucial to estimate the 
radio flux  of the whole remnant. 
We used a method based  on a color-color criteria proposed by  \citet{rr06}  
using \textit{Spitzer} data  at 3.6, 4.5, 5.8, and 8 $\mu$m
 from the Galactic Legacy Infrared Mid-Plane Survey Extraordinaire 
\citep[GLIMPSE, ][]{bc03}. The spatial resolution of the survey is about 2 arcsec.
More details can be found in Sect. \ref{sect:radioflux}.

\subsection{\x data}

The first set of observations of the SNR \rxj \citep[presented in ][]{cdg04}  covered almost the entire remnant.
 New observations of the south, east and west regions completed  the mosaic of the remnant.
With a total of 11 pointings (see Table \ref{tab:obs}) we can now provide the first full mapping of 
\rxj with \x (see Fig. \ref{fig:mosaic}).\\
To clean proton flare contamination in the event files, we built a histogram
of counts/seconds in the 10-12 keV band for the MOS cameras (12-14 keV for PN).
Then we fit a gaussian distribution upon the histogram and retain, in the observation, only the time intervals where the
count rates are within a 3$\sigma$ range \citep{pa02}. Table \ref{tab:obs} provides the list of the remaining
exposure times after flare screening. For the PN instrument only 6 pointings were used.

\subsection{HESS data}
\label{sect:hessdata}

In the HESS study (AH06) spectra were extracted  from 14 square regions 
(0.26$\degr$ length, see Fig. \ref{fig:4contrib}) covering the whole remnant.
The data used for the spectral analysis have a cut on the minimum size of image of 80 photo-electrons
resulting on a mean spatial resolution of 0.12$\degr$ (68\% containment radius).
 For the morphological analysis (radial profile) the cut used in HESS data is at 200 photo-electrons
 resulting on a better spatial  resolution of 0.08$\degr$.
The detection efficiency varies over the 5$\degr$ of the full field of view  but it
is almost constant at the scale of the remnant which is $\sim$ 1$\degr$ wide 
 (it varies by only  5\% between the centre and the edge of the object).
All the data used in our study are taken from AH06 and we did not reprocess any TeV data.
For the comparison of the spectral properties we used the \g-ray results presented in Table 2 of AH06 and 
for the comparison of radial profile we used the data presented in Fig. 16.

\section{X-ray processing}
\label{sect:xrproc}

\subsection{Mosaic construction}
   \label{sect:mosaic}

\begin{figure}
   \centering
   \includegraphics[bb= 6 152 580 640, clip,width=\columnwidth]{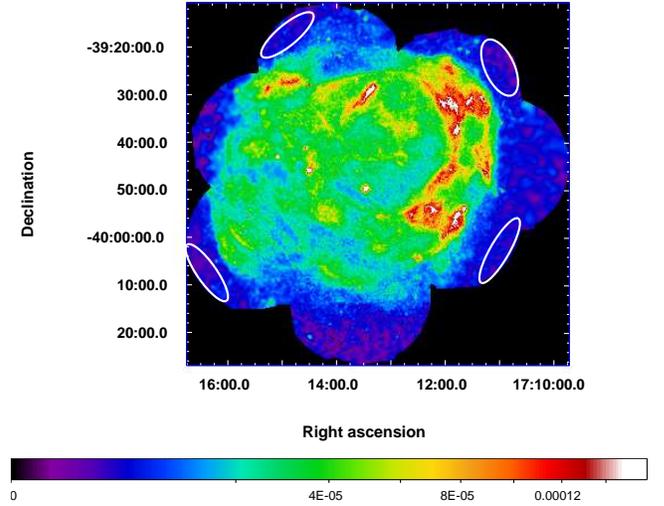}
   \caption{EPIC MOS plus PN image in the 0.5-4.5 keV band. The units are ph/cm$^{2}$/s/arcmin$^{2}$ 
and the scale is   square root.  
The image was adaptively smoothed to a signal-to-noise ratio of 10. The four ellipses show the
     regions used to estimate the local astrophysical background for the spectral analysis.}
   \label{fig:mosaic}
\end{figure}

\begin{figure*}[!t]
   \centering
   \begin{tabular}{cccc}

 { \includegraphics[bb= 116 270 467 590 , clip,width=4.5cm]{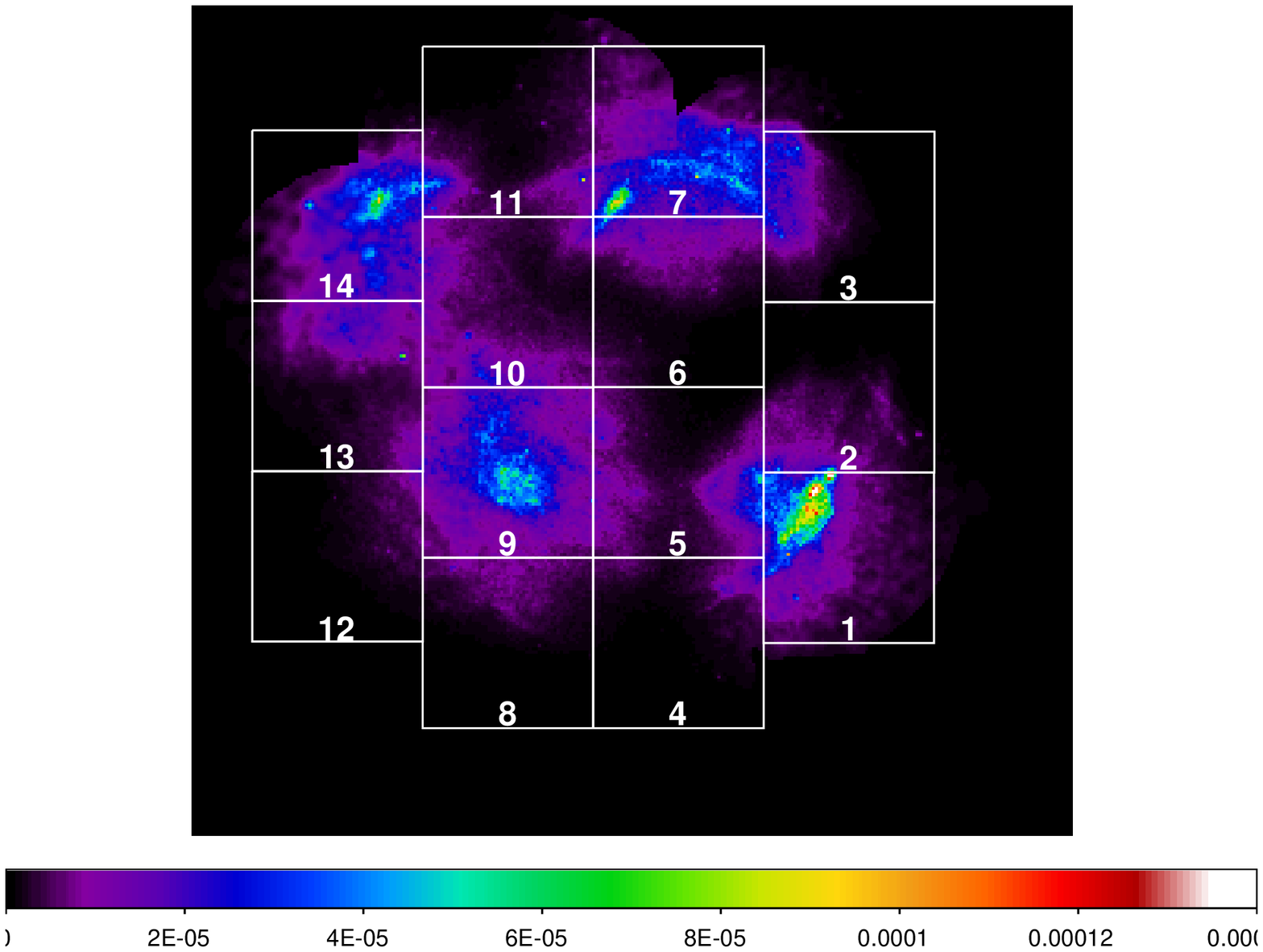}  } & \hspace{-6mm}
  { \includegraphics[bb= 116 270 467 590 , clip,width=4.5cm]{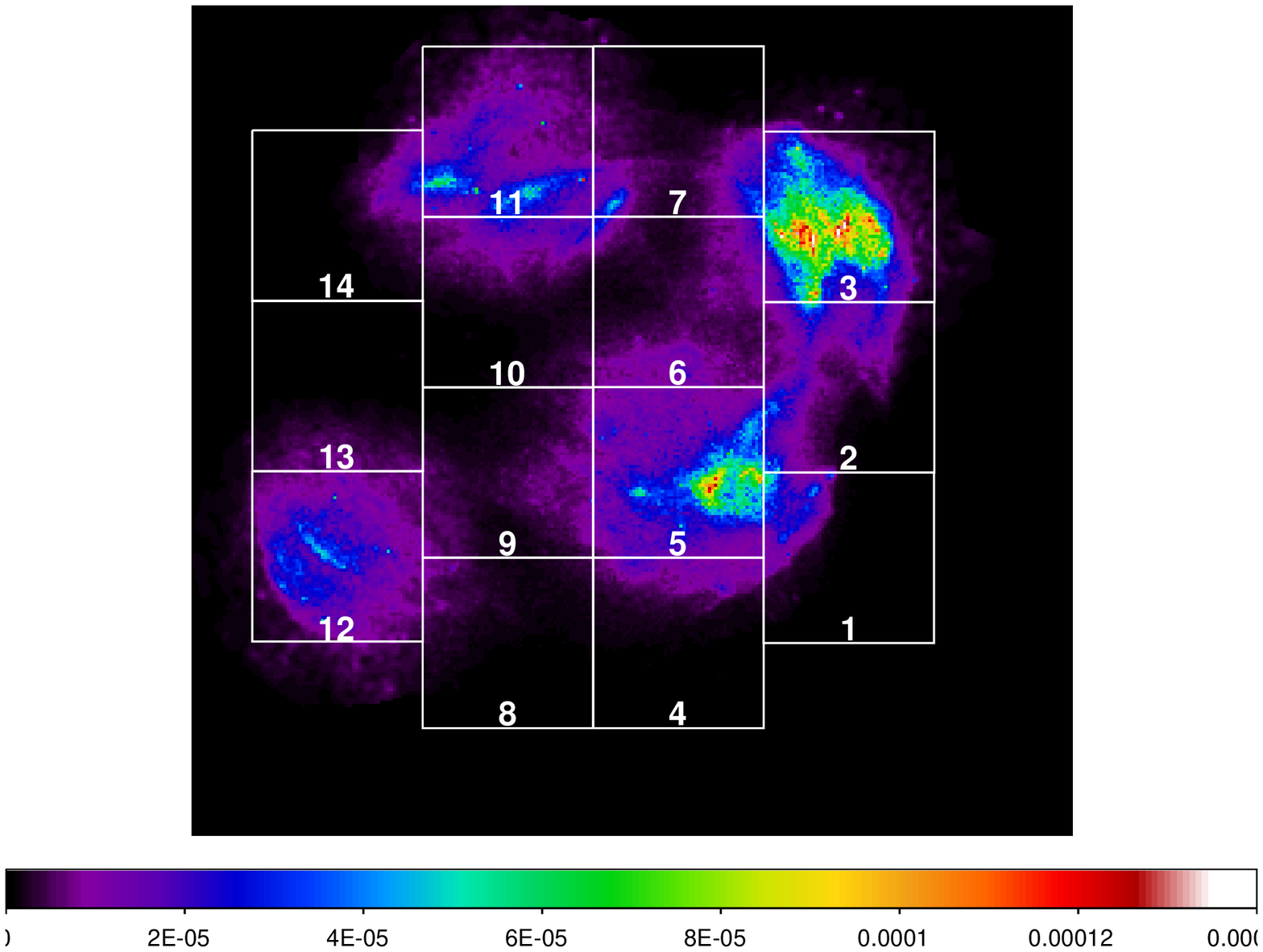} }  & \hspace{-6mm}
  { \includegraphics[bb= 116 270 467 590 , clip,width=4.5cm]{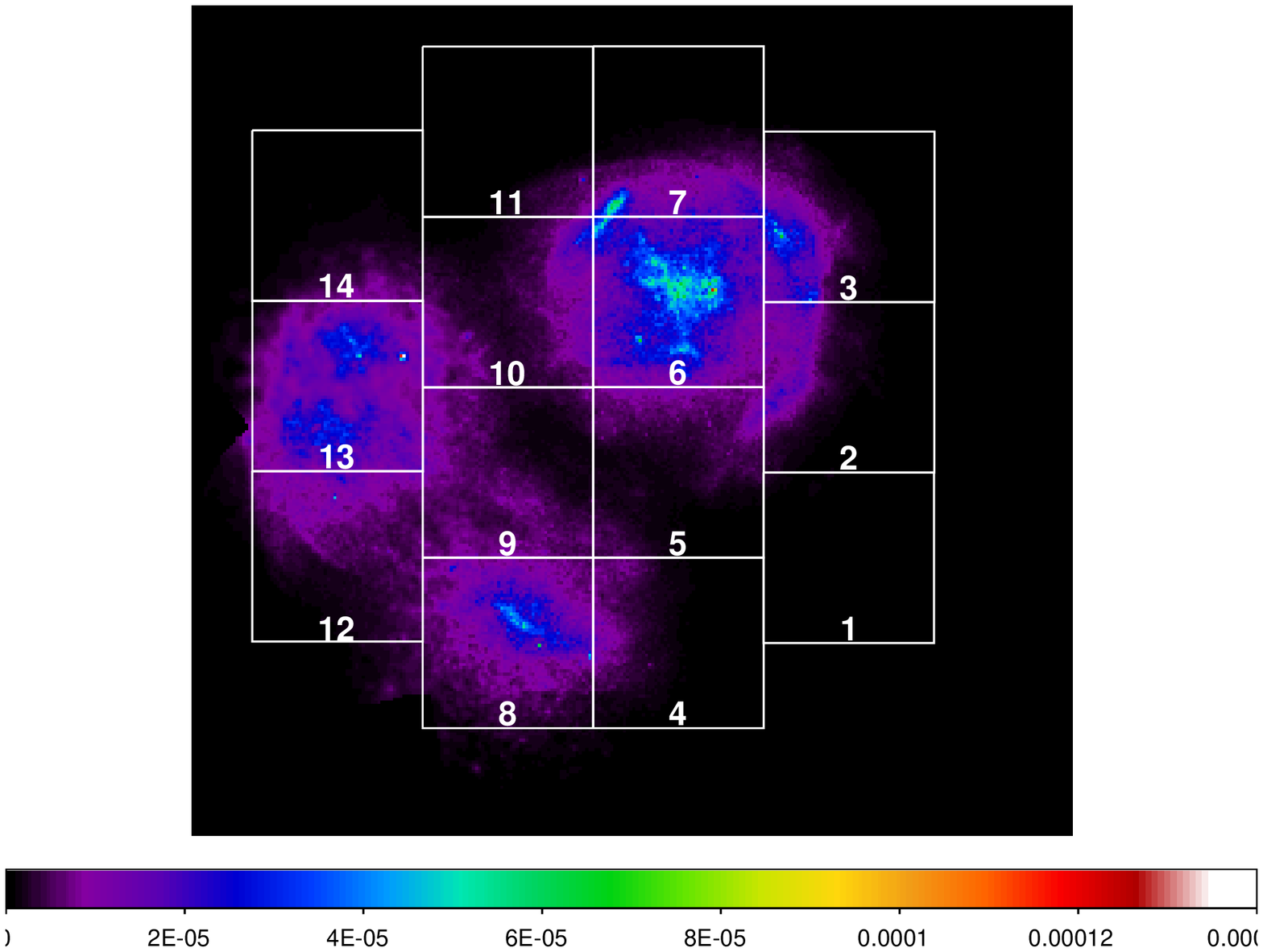} } & \hspace{-6mm}
  { \includegraphics[bb= 116 270 467 590 , clip,width=4.5cm]{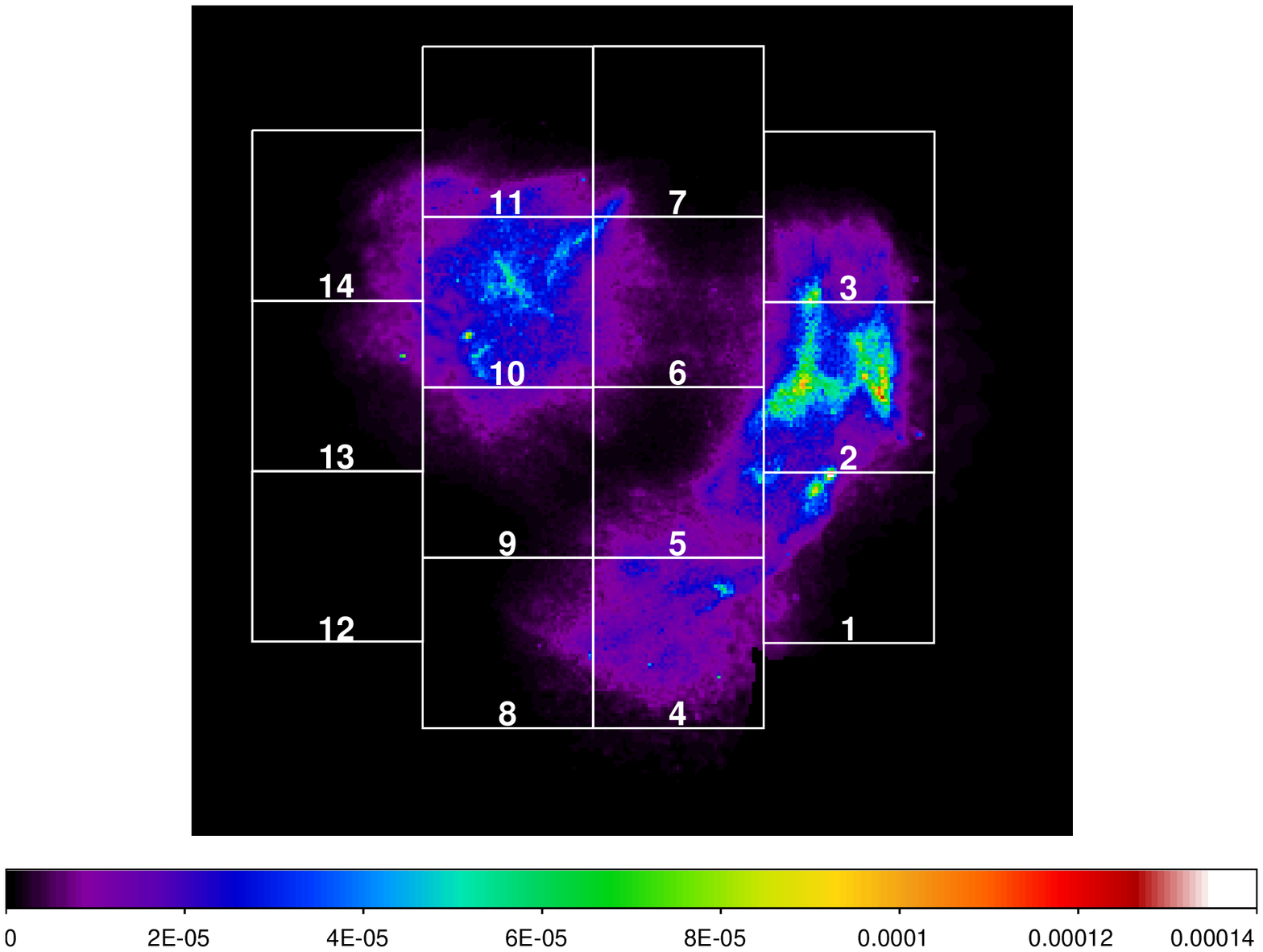} }
   \end{tabular}
   \includegraphics[bb= 36 186 640 226 , clip,width=10cm]{contribtot-26mai4.ps}

   \caption{Spatial X-ray contribution to each HESS region (as defined in \citet{ah06}).
       We see that due to the size of the PSF comparable
    with the size of the extraction regions, many events outside of the region contribute to the spectrum.
 To save space, each frame shows the contributions to three or four separate regions at once.  The linear scale is in ph/cm$^{2}$/s/arcmin$^{2}$. The regions 1 to 14 defined here are used later for the spectral analysis in Table \ref{tab:bestfit}, Fig. \ref{fig:compspec}, \ref{fig:2index} and \ref{fig:compspec2}.} 

   \label{fig:4contrib}
\end{figure*}

The mosaic is built in counts and
an adaptive smoothing is applied such that the signal-to-noise ratio is at least 10.
 The instrumental background is derived from a compilation of blank
sky observations \citep{cr07},  renormalized in the 10-12 keV energy band for the MOS cameras (12-14 keV for PN)
and subtracted from each image.  To have units in
photons/cm$^{2}$/s/pixel instead of counts/s/pixel, the exposure map of each observation
is multiplied by the average effective area in the energy band (assuming the same spectrum over the field of view).
Then a mosaic of those exposure maps is built and smoothed in the same way. 
The final image is the division of the counts mosaic by
the exposure map mosaic.\\
The resulting image which is the sum of the MOS and available PN
 data after flare screening is presented in Fig. \ref{fig:mosaic}.
The morphology of \rxj can be decomposed in two main kinds of structures :
diffuse emission present over all the
remnant and bright filaments particularly visible in the west and north of the remnant (see Fig. \ref{fig:mosaic}).
Thanks to the high sensitivity of  \x, we can clearly see the faint emission in the
recent observations of the regions south, east and north.
In particular in the northern region of the remnant we distinctly see a straight edge
that is not an artifact due to a CCD gap or any instrumental effect. Also it is not due to
an X-ray absorption along the line of sight as this straight edge  remains visible on the 4.5-7.5 keV image
 (in this energy band the absorption is weak). Simply the emission seems fainter there.
 Above that edge, we see a structure (in blue-green) that seems to be  the continuity of the shock.

The estimate of the astrophysical background is not simple in \rxj as it seems to vary around the remnant.
However in order to have a rough approximation of the background level 
we extracted the flux outside a circle of 0.56$\degr$ radius centered on the remnant
($\alpha_{\mathrm{J2000}}=$17h13m46s,  $\delta_{\mathrm{J2000}}=-39\degr44'56"$).
We then subtracted the  mean value of this flux ($4.5 \times 10^{-6}$ photons/cm$^{2}$/s/arcmin$^{2}$) 
 for the morphological study. The small structures of the SNR  are not affected by this background subtraction.

\begin{table}
 \caption{\textit{XMM-Newton} observations used in this paper. The total and good columns represent
 the exposure time before and after flare screening.}
 \label{tab:obs}
 \begin{tabular}{l|l|c c|c}

 \hline
 \hline
   &                  &     \multicolumn{3}{c}{Exposure (ks) }   \\ 
   &                 &      \multicolumn{2}{c}{MOS }  & PN \\
 ObsId & Observation Date &  Total & Good  & Good \\

 \hline

   0093670101 (NE) &  2001 September 5 & 15.3 & 1.8 & 0\\

   0093670201 (NW) &  2001 September 5 & 15.3 & 6.7  & 0\\

   0093670301 (SW) &  2001 September 8 & 15.3 & 15.2  & 10.0\\

   0093670401 (SE) &  2002 March 14 & 14.1 & 11.6  & 5.1\\

   0093670501 (CE) &  2001 March 2  & 13.8 & 13.0  & 6.5\\

   0207300201 (CE) &  2004 February 22 & 31.5 & 12.4  & 0.\\

   0203470401 (NE) &  2004 March 25 & 17.0 & 16.1   & 6.7\\

   0203470501 (NW) &  2004 March 25  & 18.0 & 13.1  & 9.7\\

   0502080101 (E) &  2007 September 15 & 34.6 & 5.8  & 0\\
   
   0502080301 (W) &  2007 October 3 & 8.9 & 2.8  & 0\\

   0551030101 (S) &  2008 September 27 & 24.9 & 24.5 & 20.8\\

 \hline
 \end{tabular}
 \end{table}

\subsection{Spectral extraction method}
   \label{sect:specmethod}

With its good spatial resolution, the \x  telescope can carry out spectral study at small scale whereas in \g-rays the
spectral analysis is done at larger scale due to the comparatively lower spatial resolution of the HESS telescopes.
To address this problem, 
we  took into account the different Point Spread Functions and the variation of the detection 
efficiency  across the field of view of the two instruments.

In the case of \x,  the detection efficiency of the MOS and PN cameras  can drop 35\%  from the
centre to the edge of our  0.26$\degr$ extraction regions.
For the HESS telescopes the detection efficiency is almost constant to this size (see Sect \ref{sect:hessdata}).
When extracting an X-ray spectrum in this large region we want all the events to contribute
with the same weight to the spectrum.
To address this problem, we used the weight method described in \cite{an01}
 where each event is corrected for its efficiency loss as a function of its position on the camera and its energy.

We also have to take into account the different size of the Point Spread Function of both instruments.
For the spectral study of this SNR, the mean spatial resolution of the HESS instrument is 0.12$\degr$
(68\% containment radius)  which is comparable to
the size of the extraction region (0.26$\degr$) whereas with the \x observatory, the spatial resolution is about 7 arcsecs.

We assumed that the Point Spread Function of the HESS telescopes is a gaussian of $\sigma=0.0795\degr$ (corresponding to the HESS 68\% containment radius of 0.12$\degr$) and that in comparison the  
Point Spread Function of \x is negligible. To decrease the spatial resolution of the X-ray data,
 we redistributed randomly the position of each X-ray event according to the gaussian probability density function.

For the purpose of spectral comparison with the \g-rays, we removed the two following very bright point-like sources in the X-ray data : 1WGA J1713.4-3949 which is argued  to be the Central Compact Object of the SNR
 \citep{sl99,ls03,cdg04} and 1WGA J1714.4-3945 which is associated with a star  \citep{pa96}.

\subsection{X-ray background spectra}
   \label{sect:bkgsub}

The instrumental background spectrum is derived from blank sky observations \citep{cr07} in the same detector area,
renormalized in the 10-12 keV energy band for the MOS cameras (12-14 keV for PN) and processed with the
 same method as  the observations (see Sect. \ref{sect:specmethod}).

The subtraction of the local astrophysical background, 
 is not a simple problem in \rxj. On the one hand, as the remnant is $\sim 1\degr$ in size and close
 to the galactic plane,
the local astrophysical emission can vary around the remnant. On the other hand,
there are only few pointings that allow us to estimate this background.
We extracted background spectra outside of the SNR in 4 regions where the statistics was sufficient 
(see the 4 ellipses  in  Fig. \ref{fig:mosaic}).
 In this astrophysical background,   no emission lines are seen
but the statistics is low. For comparing the spectral
properties of the background between the different regions,
 we modeled the spectra using a simple absorbed power law model.
The parameters of this model are similar in the northern, north-western and south-western regions i.e. :
$\textit{N}_{\mathrm{H}} \simeq 0.4 \times 10^{22}$  cm$^{-2}$, index $\simeq$ 1.9 and a normalization
$norm \simeq 1.4 \times 10^{-3}$ keV$^{-1}$cm$^{-2}$s$^{-1}$ at 1 keV 
(renormalized to the area of the large HESS extraction regions). 
Note that in the south-eastern region, the normalization parameter is 40 \% smaller.
As all the extracted background spectra have low statistics, we decided to use a power law model with the parameters listed above
for the astrophysical background instead of a spectrum.
 We used the same $\textit{N}_{\mathrm{H}}$, index and normalization parameters that
 of the northern, north-western and south-western regions for all
the extraction regions. The impact of such a choice for the south-eastern region, 
where the normalization is smaller, is discussed later in Sect. \ref{sect:specmod}.

\subsection{Spectral modeling}
   \label{sect:specmod}

\begin{figure}[!h]
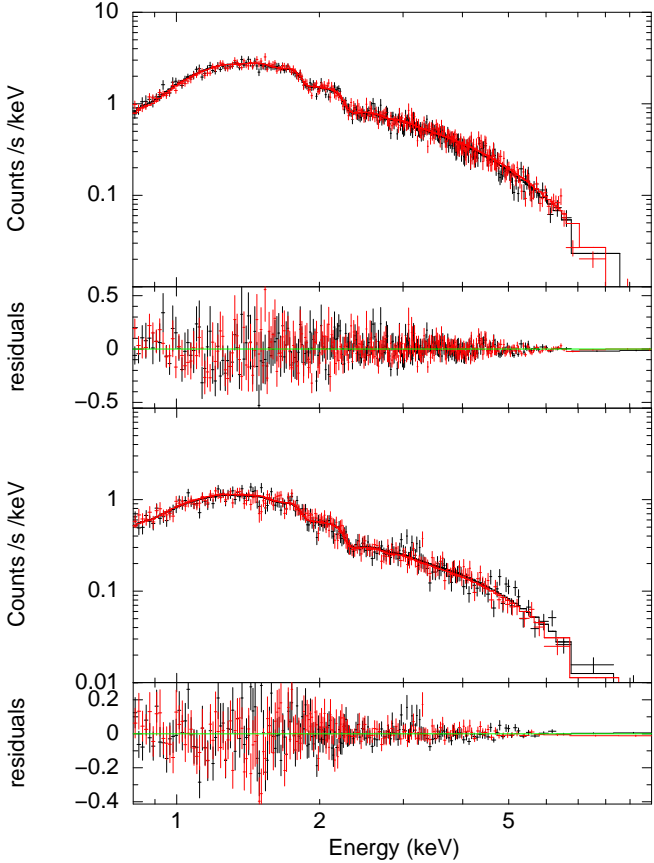


   \centering

  \includegraphics[angle=-90,bb= 56 -20 514 705, clip,scale=0.34]{region3.ps} \vspace{-2.62mm} \\
 \includegraphics[angle=-90,bb=  56 -14 575 705, clip,scale=0.34]{region12.ps}

   \caption{Best-fit X-ray spectrum from region 3 (\textit{Top} panel) and region 12 (\textit{Bottom} panel)
    with an absorbed power law model  (MOS1 spectrum is in black and MOS2 in red). 
The region 3, located on the north-west,  is the brightest  region of the remnant. It has a steep spectrum (2.35) and has a high absorption ($0.72 \times 10^{22}$cm$^{-2}$).
 On the  south-east of the remnant, 
   region 12 is faint, has a weak absorption ($0.50 \times 10^{22}$cm$^{-2}$) and has a harder spectrum  (2.17).}

     \label{fig:2spectra}
  \end{figure}

The X-ray emission of the remnant is dominated by a non-thermal continuum and no emission lines have
 been detected so far \citep{kk97,sl99,pa03,cdg04}.
 In our study, all spectra are well described by an absorbed power law model as is
 illustrated  in Fig. \ref{fig:2spectra}.
In some observations, the PN instrument exposure time was null after flare screening. 
To have an homogeneous coverage of the remnant we kept only the MOS data.
Whenever it was possible we fitted independently, as a test, the spectrum extracted from the MOS1\&2 and PN instruments and compared the
resulting best-fit parameters. They agreed within the statistical errors bars. \\
All the data was fitted using unbinned spectra with the C-statistic implemented in Xspec (v12.3.1).
Binning was used for graphical purposes only and fixed at 3$\sigma$ for all spectra. We have fitted 
 all the data from 0.8 keV to 10 keV.
The best-fit parameters for the HESS regions are listed in Table \ref{tab:bestfit}.
As discussed in Sect. \ref{sect:bkgsub}, the local astrophysical background in the south-eastern region
 seems weaker than the average background used for all the regions.
 Therefore, we studied the impact of those different backgrounds for the  faintest region 
(the most dependant to background subtraction) which is located in the SE : region 12. 
When using a value of the normalization parameter
40 \% smaller for the background model, the new best-fit parameters do not vary much.
While the hydrogen column and the index change within the errors bars, the integrated X-ray flux 
varies by less than 10\%. As the variations are small and do not impact the results of our study,
 we kept the same astrophysical background for all the regions.

\subsection{Radial profile method}
   \label{sect:radmethod}

\begin{figure}[!h]

   \centering
 \includegraphics[width=8.cm]{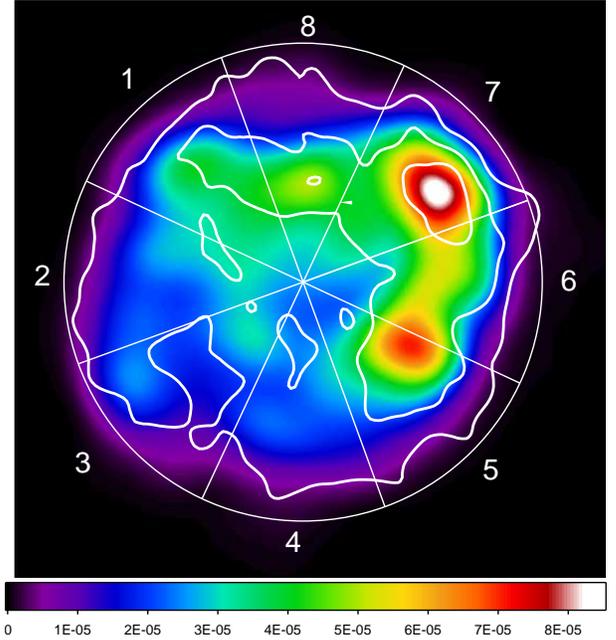}
  
 \caption{Same X-ray mosaic as Fig. \ref{fig:mosaic} smoothed  to match the
  Point Spread Function of the HESS telescopes. The scale is linear and units are in ph/cm$^{2}$/s/arcmin$^{2}$.
 Overlaid are the \g-ray contour excess (spaced at 30, 60 and 90 counts) from Fig. 7 of \citet{ah06}.
    The sectors used for the radial profile comparison of Fig. \ref{fig:radprofiles} are also drawn.}
   \label{fig:regioncam}
\end{figure}

As we now have a full coverage of the remnant, we
can properly compare the radial profiles of the emission of the remnant in X- and \g-rays.
 To match the HESS spatial resolution we smoothed the X-ray image with a gaussian of $\sigma$=0.053$\degr$
 (corresponding to r$_{\mathrm{68\%}}$=0.08$\degr$).
 We then extracted a flux per unit solid angle as function of the distance to the center
 on the smoothed X-ray  images (1-2 keV and 2-4.5 keV band)  in eight sectors as shown in Fig. \ref{fig:regioncam}.
 To compare  the radial
 profiles at both wavelengths,  the X-ray  profiles were scaled by a unique normalisation factor
 calculated as the ratio of the
 total number of counts in \g-rays over the total flux in X-rays on the whole remnant.
 In the last two bins of some X-ray profiles,
  there is not  enough coverage to estimate the flux. Those bins are removed from the profile.
As in the spectral study, the two brightest point-like sources (see Sect. \ref{sect:specmethod})
were removed from the X-ray images. The resulting radial profiles are shown in Fig. \ref{fig:radprofiles}
and discussed in Sect. \ref{sect:radprofresults}.

\section{Results}
   \label{sect:results}

\subsection{Radio flux}
\label{sect:radioflux}

An accurate estimate of the integrated radio flux density is critical
 to investigate the mechanisms responsible for the high-energy
 emission.  In this work we re-analyze the available radio data at 1.4 GHz
 following two different paths to estimate the radio flux density of
 RX J1713.7-3946. 

\begin{figure*}[t]
\centering
%

\includegraphics[width=8.5cm]{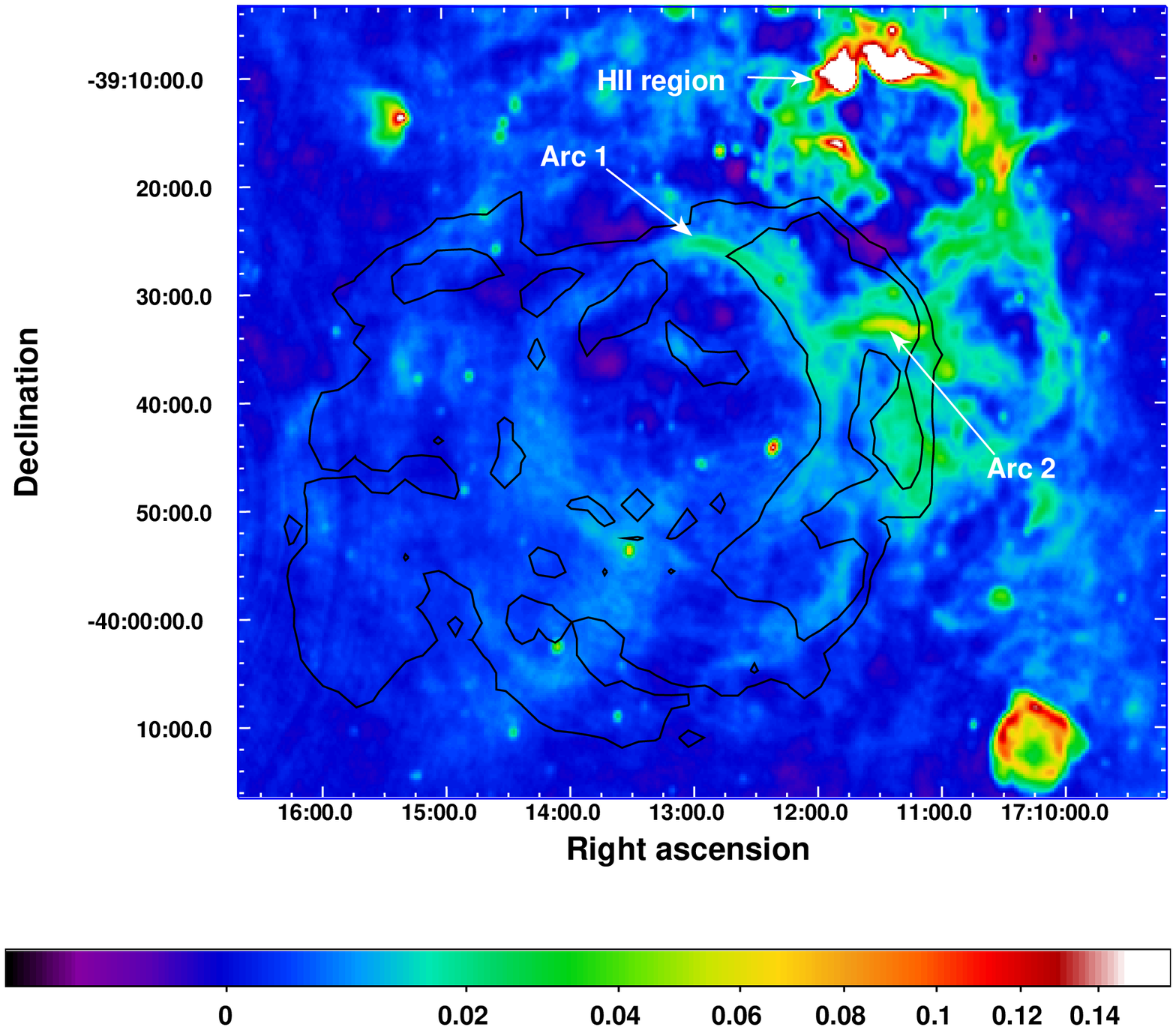} \hspace{0.4cm}
\includegraphics[width=8.5cm]{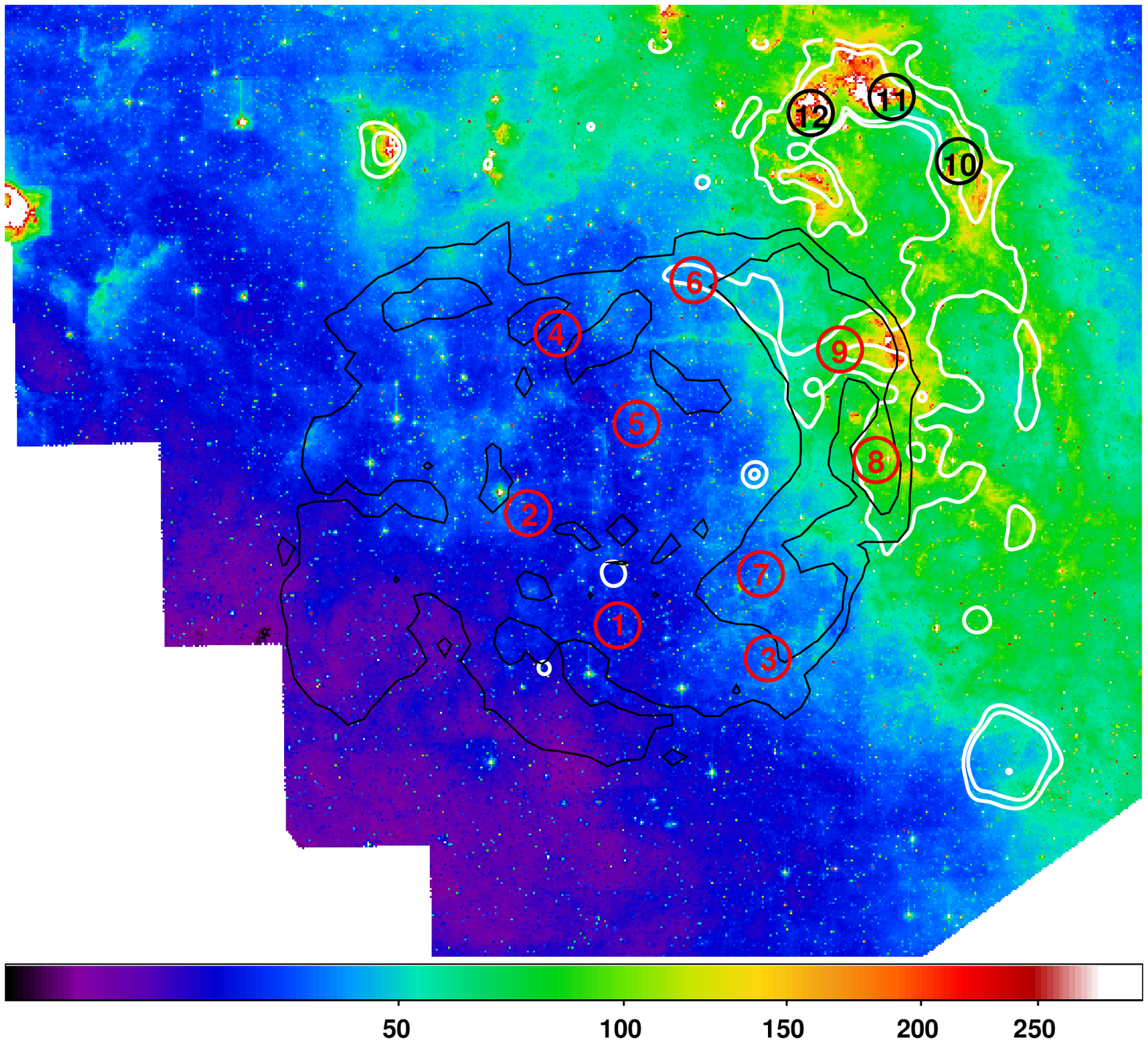} \\
\vspace{0.cm}

\caption{{\it Left} : Radio image of the region of \rxj at 1.4 GHz
\citep[from][]{ls04}. The scale is square root and  units are in Jy beam$^{-1}$. {\it Right} : {\it Spitzer} image at
$8 \, \mu$m from the GLIMPSE survey. The X-ray contours are represented in black in both images
and the radio contours in white. The regions used for the color-color plot of Fig. \ref{fig:ircolors}
 are labeled here. The scale is square root and  units are in MJy sr$^{-1}$.}
\label{fig:radioir}
\end{figure*}

Since the largest well imaged structure
for the ATCA observations is 25 arcmin in size, to recover information of
structures larger than this size, the combination with single dish
observations is required. However, in this case, it was not possible to
complete the procedure due to the fact that the
overlap annulus of the interferometric and single antenna
data set\footnote{The only public single dish
data available at this frequency come from  the Survey of the South
Celestial Hemisphere carried out with the 30 m dish of the Argentine
Institute of Radio Astronomy \citep[IAR][]{tr01}  (HPBW$\sim$34 arcmin at
1.4 GHz)}  in the uv space is too small to produce a reliable image. 
We
therefore used the single dish data at 1.4 GHz to estimate the integrated
flux density within the area covered by the X-ray emission associated with
the SNR (outer contour depicted in Fig. \ref{fig:radioir}). This estimate provides an
upper limit for the flux density of about 26 Jy with an uncertainty of the order of 10\%. In addition we
 integrated over an interferometric image based on ATCA data within
 the same outer contour. 
In this case, since the largest well imaged
 structure at this frequency is about 25 arcmin, we corrected ``by
 hand'' adding the minimum flux density required to fill in the few
 negatives remaining in the image. 
The surface brightness corresponding to this addition was below $7.4\times10^{-24}$ W m$^{-2}$ Hz${-1}$ sr$^{-1}$. The off-source rms-noise is about 0.7 mJy/beam and the constant value
added to fill in the few negatives is ~ 1.2 mJy/beam.
Such correction amounts less than  1\% of the total estimated flux density.

\begin{figure}
\centering
\includegraphics[width=\columnwidth]{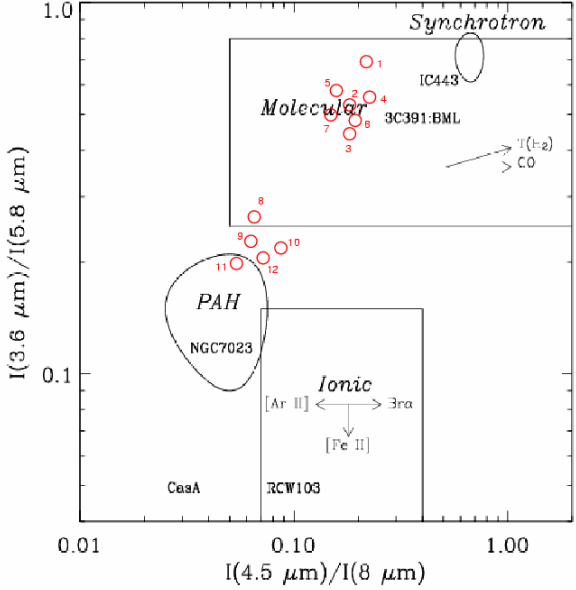} 
\caption{Infrared color-color diagram for the regions defined in Fig. \ref{fig:radioir} (\textit{Right}) overlaid
on Fig. 2 of  \citet{rr06}.  Two distincts groups can be seen. The emission from the regions 8 to 12 is
 compatible with PAH origin including our region of interest Arc 2 (labeled as region 9 here).
Therefore the radio emission from Arc 2 is likely to be thermal and not related with the remnant.}
\label{fig:ircolors}
\end{figure}

Whether the radio  emission from Arc 2 (see Fig. \ref{fig:radioir}) is thermal or non-thermal in nature 
is an important issue as it is one the brightest features within the X-ray contours.
In order to investigate the connection of Arc 2, with the SNR, we used 
mid-infrared observations.  The right-hand panel of Fig.  \ref{fig:radioir}
shows {\it Spitzer} $8 \, \mu$m mid-infrared emission in the direction of
RX J1713.7-3946.  Particularly, it can be seen intense infrared
emission at the location of the HII region G347.611+0.204. The infrared
emission also evidences good morphological correlation with the radio
Arc 2 suggesting a thermal origin for this feature. To investigate
the nature of Arc 2 we applied the color-color criteria proposed
by  \citet{rr06} based on {\it Spitzer} data obtained at 3.6,
4.5, 5.8, and 8 $\mu$m, finding that Arc 2 has color
characteristics compatible with polycyclic aromatic hydrocarbons
(PAHs) origin (see Fig. \ref{fig:ircolors}). 
Besides we applied HI
absorption techniques based on data extracted from the Southern
Galactic Plane Survey \citep[SGPS; ][]{mc05} together with a
flat rotation model for our Galaxy (assuming as solar parameters
$R_{\odot}$ = $7.6~ \pm ~0.3$ kpc and $\Theta_{\odot} =214 \pm 7$
\kms). From this study we conclude that the most
distant HI absorption feature for the Arc 2 is at -120 km/s, which
corresponds to a near distance of $\sim$ 6.7 kpc, placing in principle
this thermal arc beyond RX J1713.7-3946.
With this information, to estimate the associated radio flux density we
subtracted the contribution from all overlapping radio point sources
(likely to be extragalactic and/or compact HII regions) and from Arc
2 (about 1.5 Jy) because it is likely to be unrelated with the remnant, estimating a lower limit for
the total flux density of $\sim$ 22 Jy. Therefore, we 
 conclude that the associated radio flux density at 1.4 GHz
is  between 22 and 26 Jy.

\subsection{Global X-ray flux}
\label{sect:xflux}

We  extracted a MOS spectrum from the whole remnant without degrading the spatial resolution of the data. 
The best-fit parameters (with background subtraction as described in Sect. \ref{sect:bkgsub}) are 
$\textit{N}_{\mathrm{H}} = 0.66 \pm 0.01  \times 10^{22}$  cm$^{-2}$, index=2.37 $\pm$  0.01,
 an absorbed 1-10 keV flux of 3.95 $\pm 0.03  \times 10^{-10}$ erg cm$^{-2}$s$^{-1}$
 and a non absorbed flux in the same band of 5.23 $\pm 0.04  \times 10^{-10}$ erg cm$^{-2}$s$^{-1}$.
Our main source of uncertainty is  the absolute calibration of \x which is known with a precision of 10\%
rather than the very small statistical error (less than 1\%).
Another flux estimate is obtained with a srcut synchrotron model using the radio flux upper limit derived in Sect. \ref{sect:radioflux}. The
non absorbed flux derived is 5.49 $\pm 0.05  \times 10^{-10}$ erg cm$^{-2}$s$^{-1}$, very similar to the one obtained with a power law model. In the \textit{srcut} model, the radio spectral index was fixed to 0.6 and a break frequency of 4.2  $\times 10^{17}$ Hz was derived. This value, averaged across the remnant, is similar to the maximum break frequency reached in the bright limbs of SN 1006 \citep{rb04}. \\

We note that the 1-10 keV non absorbed flux obtained by \citet{tu08} with \textit{Suzaku} is 47\% higher
(7.65  $\times 10^{-10}$  erg cm$^{-2}$s$^{-1}$)  than what we have found. However
 the absorption and index of their spectrum ($0.79 \times 10^{22}$  cm$^{-2}$ and 2.39 respectively) 
are in agreement with our parameters. 
It is important to note that whereas our spectrum is extracted directly on the whole remnant, 
the \textit{Suzaku} spectrum is the sum of spectra from 10 particular regions  scaled up to the whole
remnant assuming the surface brightness from the \textit{ASCA} image \citep[Sect. 3.3 of ][]{tu08}.
In our spectrum, the point sources were removed but their contribution to the total flux is weak 
(less than 1\% for the Central Compact Object). We have cross checked the value of our absorbed global flux 
derived from the spectrum (in the 0.5-4.5 keV energy band)
to the one derived from our mosaiced image. Both fluxes agree  within 5\%. \\

\subsection{X- and \g-ray comparison}

\subsubsection{Spectral results}
   \label{sect:specresults}

\begin{table}
\begin{minipage}{\columnwidth}
 \caption{Best-fit X-ray parameters obtained with an absorbed power law model for the HESS regions defined in
Fig. \ref{fig:4contrib}. 
The X-ray flux is
 integrated in the 1-10 keV band.}
 \label{tab:bestfit}
 \centering
 \begin{tabular}{c|c c c }

 \hline
  Region       &   $\textit{N}_{\mathrm{H}}$   &   Index    &   Integrated flux\footnote{Corrected for absorption}   \\
               &           $10^{22}$cm$^{-2}$                    &            &  ($10^{-2}$cm$^{-2}$s$^{-1}$ ) \\
\hline

1 & 0.79 & 2.43 (2.42-2.44) & 0.91 (0.90-0.92)  \\
2 & 0.62 & 2.35 (2.34-2.35) & 1.58 (1.57-1.59)  \\
3 & 0.72 & 2.35 (2.35-2.36) & 1.74 (1.73-1.75)  \\
4 & 0.76 & 2.37 (2.37-2.39) & 0.76 (0.75-0.77)  \\
5 & 0.59 & 2.32 (2.31-2.32) & 1.42 (1.41-1.43)  \\
6 & 0.56 & 2.44 (2.43-2.44) & 1.44 (1.43-1.45)  \\
7 & 0.67 & 2.41 (2.40-2.42) & 0.89 (0.87-0.90)  \\
8 & 0.64 & 2.23 (2.22-2.24) & 0.58 (0.57-0.58)  \\
9 & 0.48 & 2.22 (2.22-2.23) & 0.90 (0.89-0.91)  \\
10 & 0.46 & 2.32 (2.31-2.33) & 1.15 (1.14-1.16)  \\
11 & 0.63 & 2.35 (2.34-2.36) & 0.72 (0.72-0.74)  \\
12 & 0.50 & 2.17 (2.15-2.18) & 0.51 (0.50-0.52)  \\
13 & 0.51 & 2.31 (2.30-2.32) & 0.69 (0.68-0.70)  \\
14 & 0.57 & 2.28 (2.26-2.29) & 0.70 (0.68-0.70)  \\

 \end{tabular}
\end{minipage}
 \end{table}

\begin{figure}
   \centering
 { \includegraphics[width=\columnwidth]{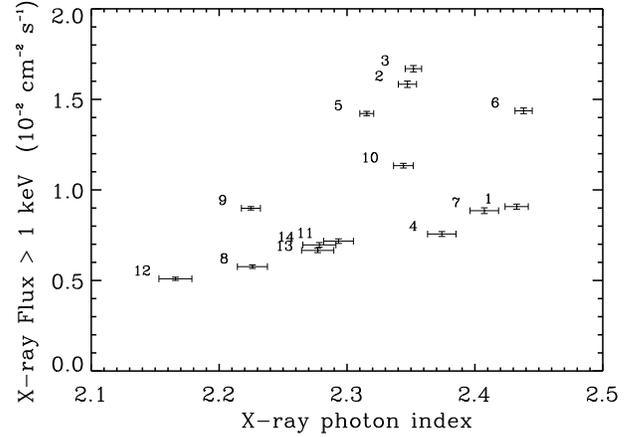} } \\
   \caption{Correlation plot between the X-ray photon index and the X-ray integrated flux
   (in the 1-10 keV band). The parameters shown here are the average of MOS1 and MOS2 best-fit parameters. 
 The label over each point corresponds to HESS regions as defined in Fig. \ref{fig:4contrib}.}
 \label{fig:compspec}
\end{figure}

\begin{figure}
   \centering
   \includegraphics[trim= 4mm 3mm 19mm 2mm,clip,width=\columnwidth]{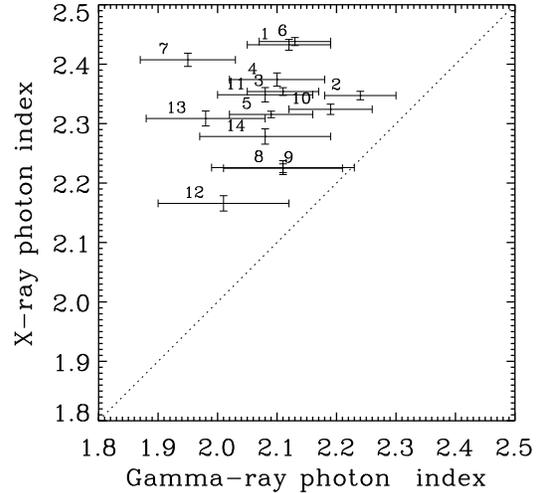}
   \caption{X-ray photon index against \g-ray photon index.
 A systematic error of 0.1 is to be added to the \g-ray
   photon index. The dashed line is the bisector.}
   \label{fig:2index}
\end{figure}

\begin{figure}
   \centering

 { \includegraphics[trim= 5mm 3mm 12mm 3mm,clip,width=\columnwidth]{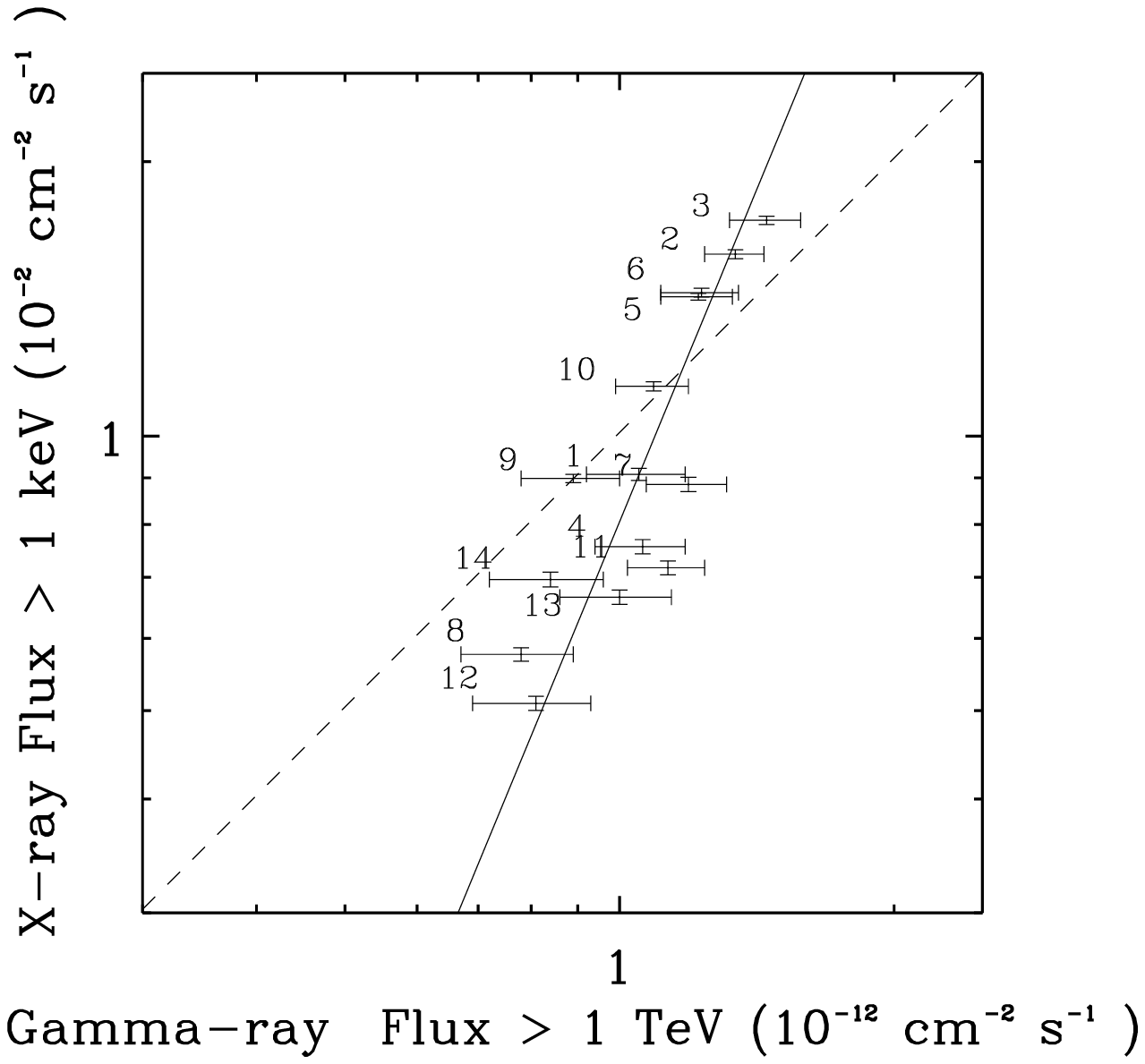} }
   \caption{Log-log correlation plot between the HESS \g-ray  integrated flux (1-10 TeV band) and the
  \x X-ray   integrated flux (1-10 keV band, using MOS1$\&$2 data). 
       We can see that the correlation is best-fitted with a non-linear function
  $F_\nu^{\rm X}=0.81 \pm 0.09 \, (F_\nu^{\rm \gamma }) ^{2.41 \pm 0.55}$  (solid line). 
The  best-fit linear function  $F_\nu^{\rm X}=1.01 \pm 0.04 \, F_\nu^{\rm \gamma }$  is also represented (dashed line).}
 \label{fig:compspec2}

\end{figure}

The best-fit parameters of the X-ray spectral modeling of the 14 regions are given  in Table \ref{tab:bestfit}.
The large variation of photon index  ($1.9<\Gamma<2.6$) seen in X-rays
 when using small extraction regions  in \cite{cdg04} have
 largely been reduced here with larger extraction regions and a degraded spatial resolution for the X-ray data
 ($2.2<\Gamma<2.4$, see Fig. \ref{fig:compspec}).
The comparison of  the X- and \g-ray photon index (Fig. \ref{fig:2index}) shows no significant correlation .
The distribution of the photon index in X- and \g-rays has a mean value of 2.32 and 2.09 respectively and a
 standard deviation of 0.075 and 0.073. Whereas the dispersion of the photon index at both energies is the same,
 the X-ray index is slightly higher than the \g-ray one.
However, there is a systematic error on the \g-ray
photon index of 0.1 (AH06) that is to be added to the Fig.\ref{fig:2index}.
Whereas   the variations of the photon index are small, there are significant variations
 in the flux from the 14 large regions.
Fig.  \ref{fig:compspec2} shows a good correlation between
the X-ray integrated flux (1-10 keV band)
and the  \g-ray integrated flux (1-10 TeV band). 
We note that the bright regions are brighter in X-rays than in \g-rays.
Such a behaviour has also been observed by \citet{tu08} with  \textit{Suzaku}.
However whereas they interpreted this as a linear correlation with some X-ray deviation, we 
interpret this as a non-linear correlation.
We  measured this non-linear correlation  placing the X-ray flux
 in the X axis (as the error bars on the X-ray flux are smaller than those in \g-rays). 
This measured slope is then inverted to obtain 
${\rm d} \log F_\nu^{\rm X} / {\rm d} \log F_\nu^{\rm \gamma }$.
The  best-fit power law function $F_\nu^{\rm X}=0.81 \pm 0.09 \, (F_\nu^{\rm \gamma }) ^{2.41 \pm 0.55}$ 
 is statistically preferred ($\chi^{2}/d.o.f.$ = 9.28/12) over the
best-fit linear function  $F_\nu^{\rm X}=1.01 \pm 0.04 \, F_\nu^{\rm \gamma }$  ($\chi^{2}/d.o.f.$ = 61.06/13). 
We note that the fit with a linear function $F_\nu^{\rm X}=aF_\nu^{\rm \gamma } - b$ (bestfit parameters : $a=2.26$ and $b=1.43$)
 gives similar results in term of $\chi^{2}$ ($\chi^{2}/d.o.f.$ = 9.38/12) than the non-linear fit 
but is not physically understandable since both images are already background subtracted.

 \subsubsection{Morphological results}
\label{sect:radprofresults}

The comparison of the radial profiles  allows us to investigate the extent of each emission as well as to
 localize their respective peaks.
Such a study has already been carried out by AH06 by comparing the ASCA and HESS data.
However as the coverage with ASCA did not always reach the boundaries of the SNR it was not possible to compare
the extension of the remnant in both wavelengths.

\begin{figure*}
   \centering
   \begin{tabular}{cccc}

 { \includegraphics[bb= 100  439 521 700, clip,scale=0.35]{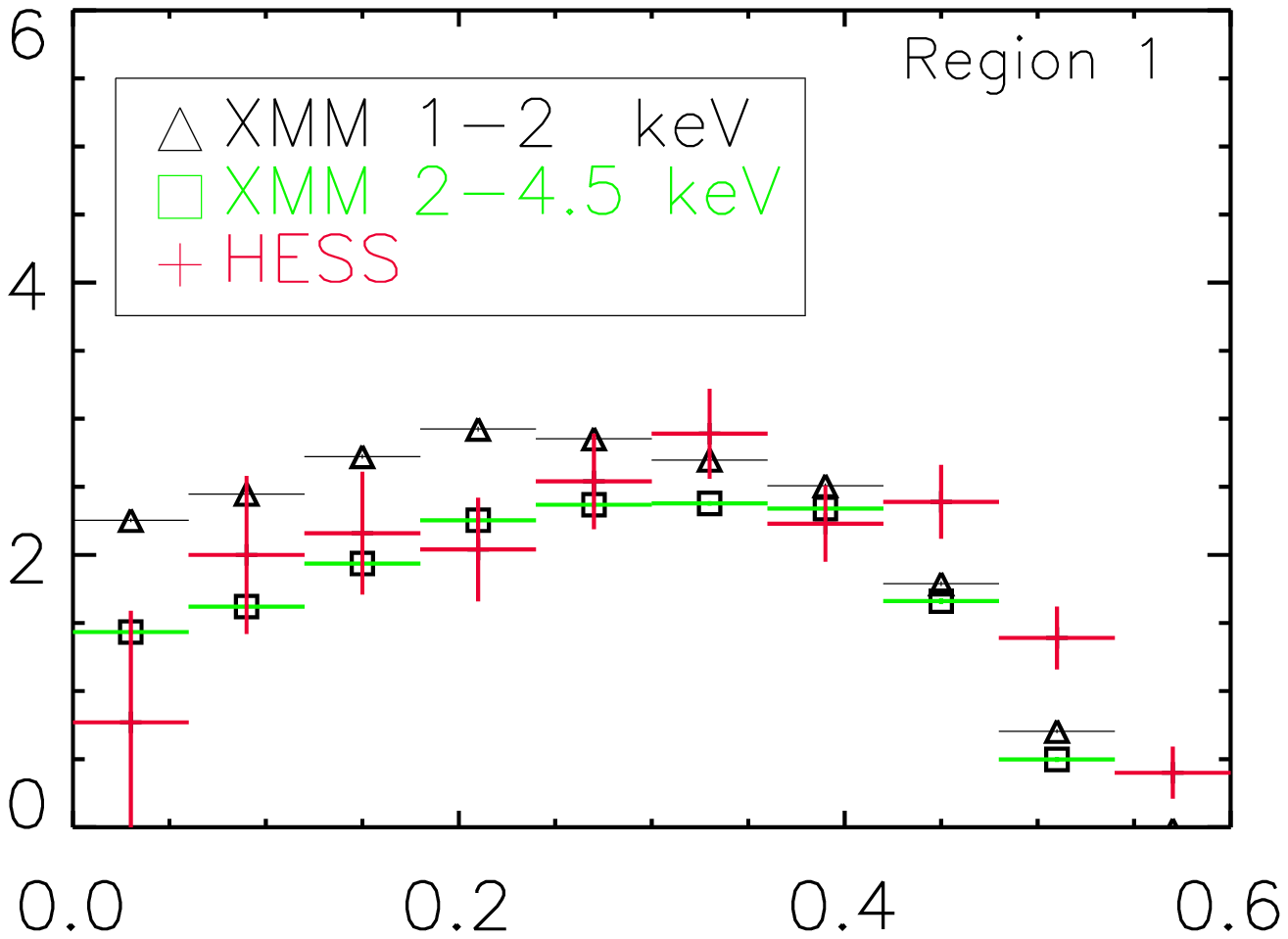} } & \hspace{-6.9mm}
 { \includegraphics[bb= 178  439 521 700, clip,scale=0.35]{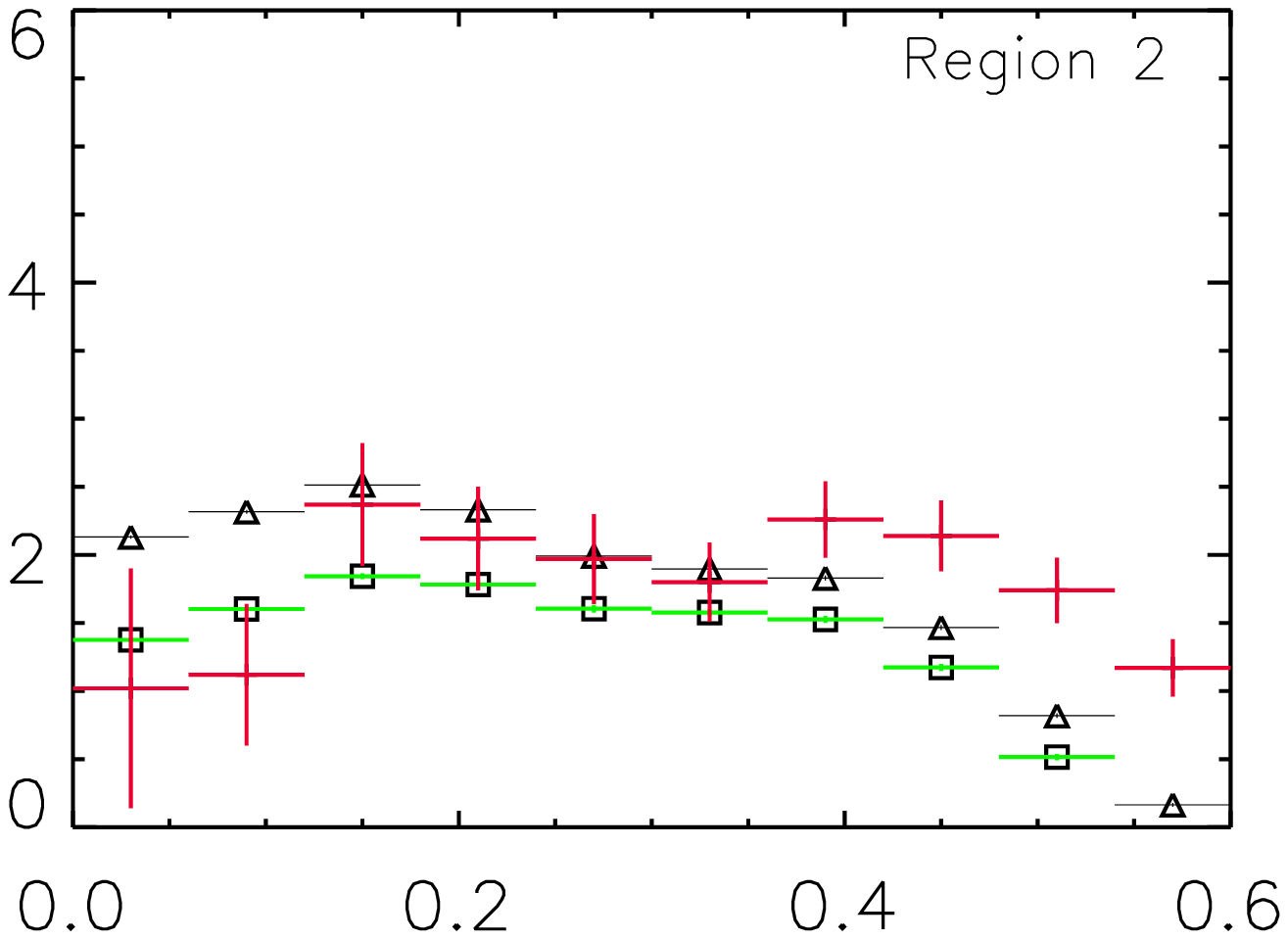} } & \hspace{-6.9mm}
 { \includegraphics[bb= 178  439 521 700, clip,scale=0.35]{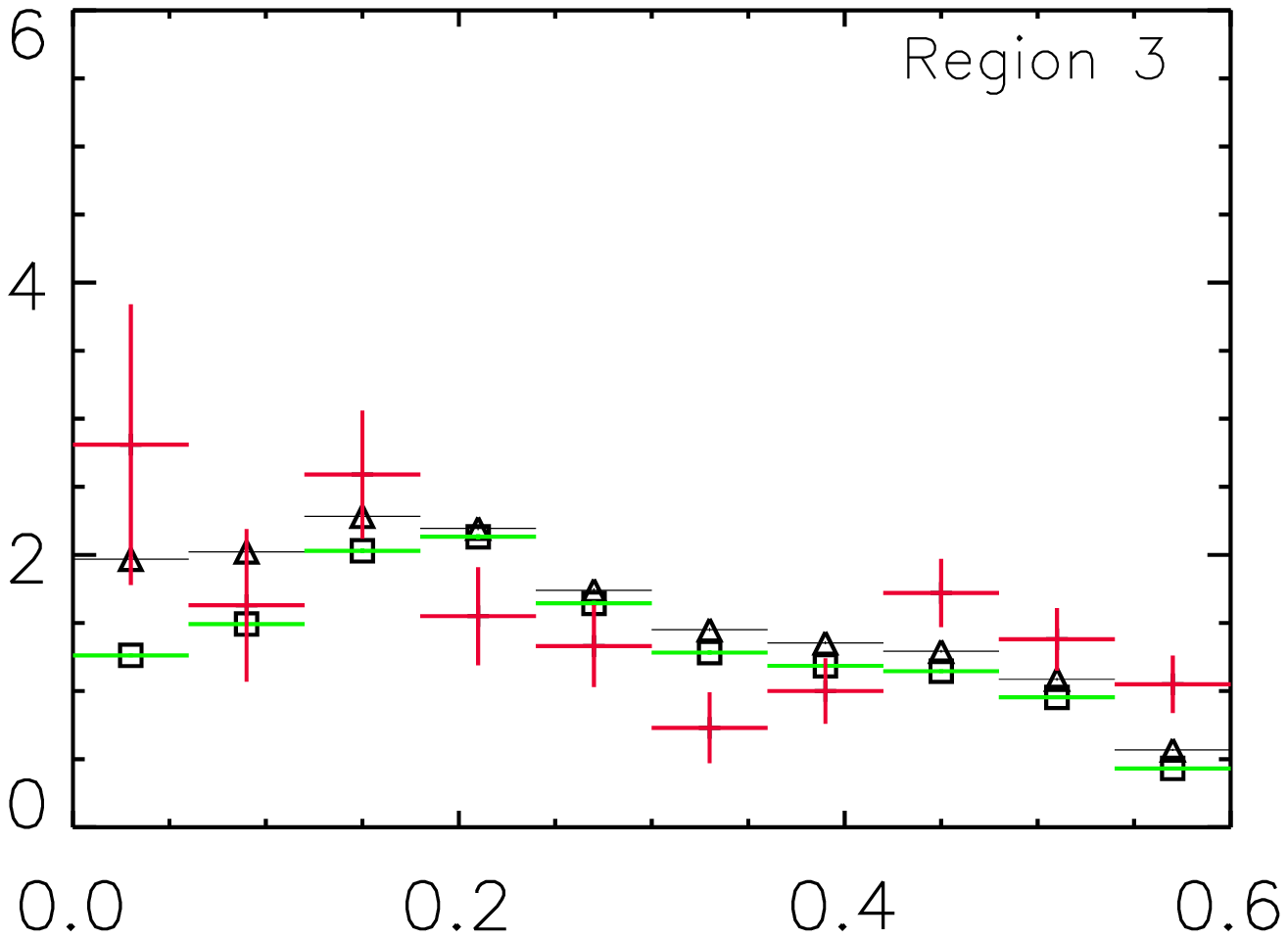} } & \hspace{-6.9mm}
 { \includegraphics[bb= 178  439 521 700, clip,scale=0.35]{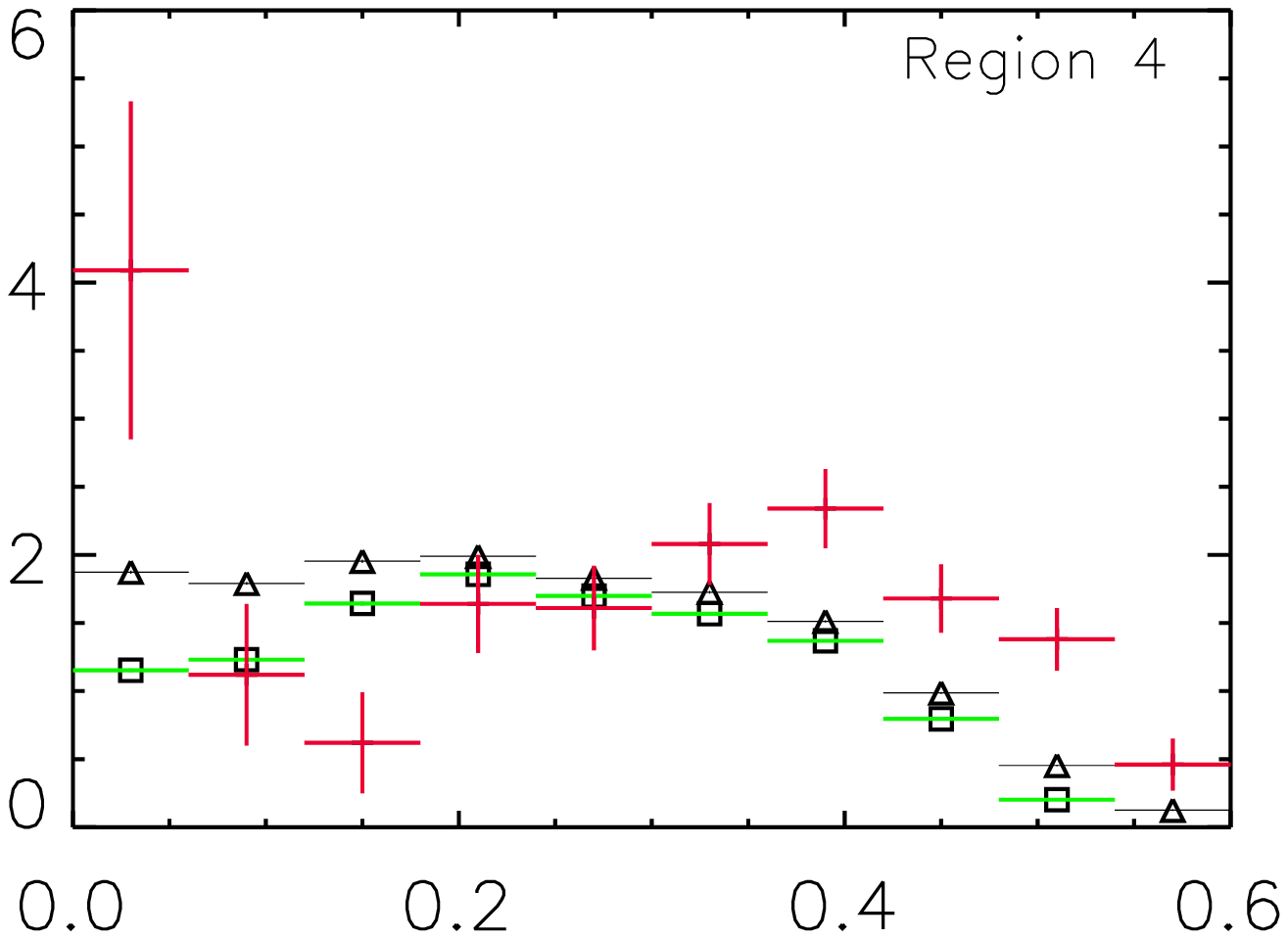} } \vspace{-3.67mm}  \\

 { \includegraphics[bb= 100  370 521 700, clip,scale=0.35]{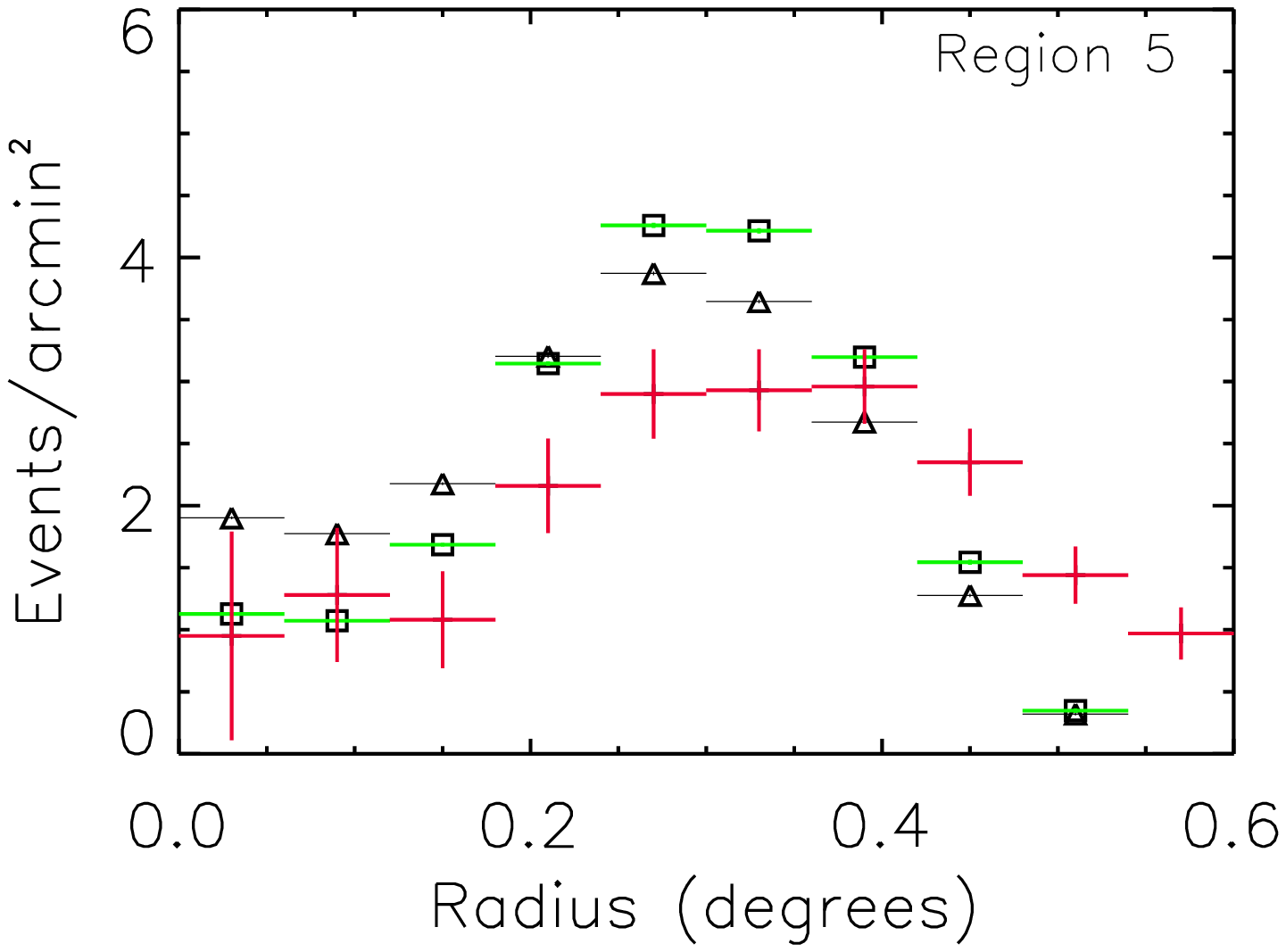} } & \hspace{-6.9mm}
 { \includegraphics[bb= 178  370 521 700, clip,scale=0.35]{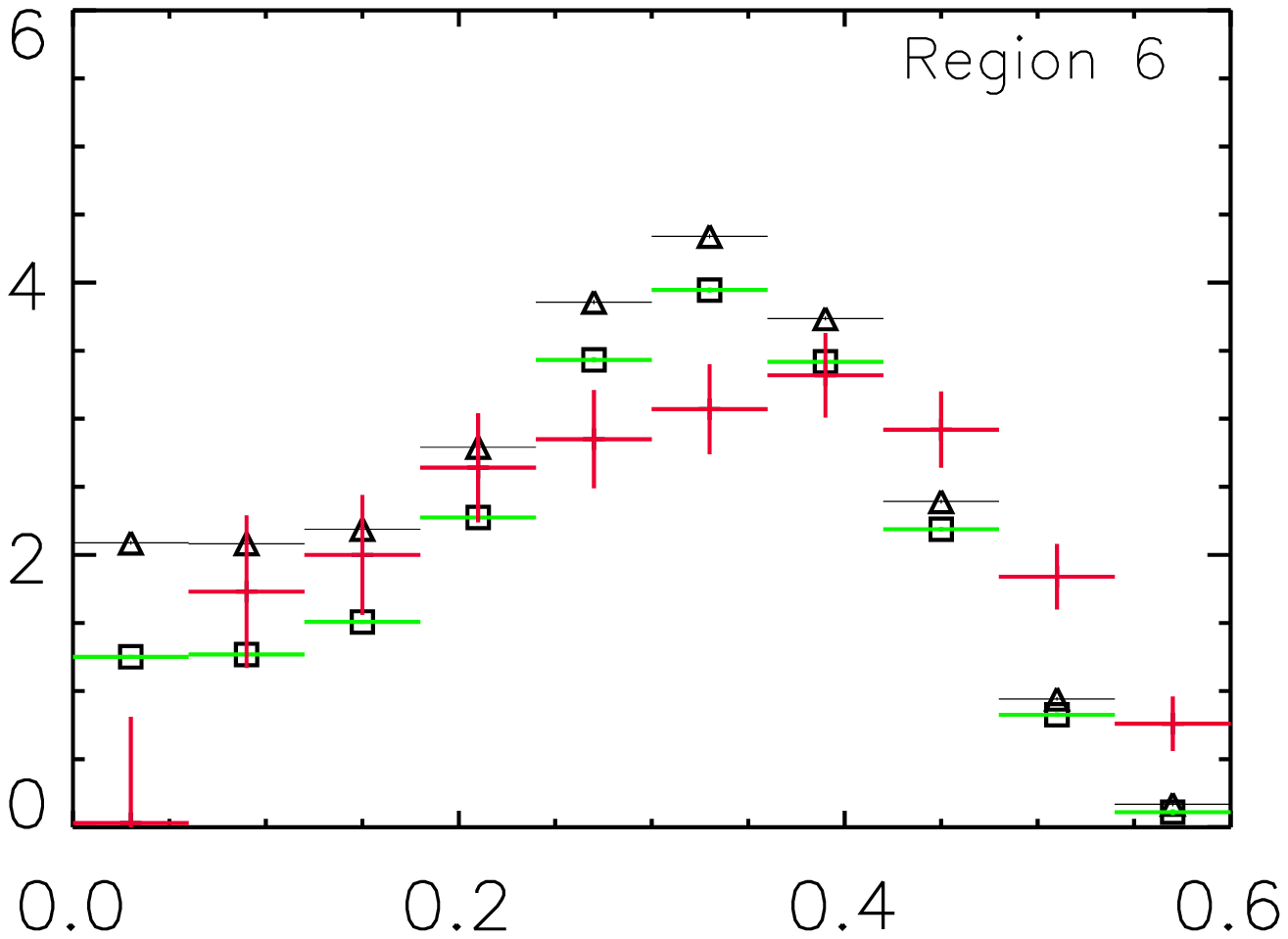} } & \hspace{-6.9mm}
 { \includegraphics[bb= 178  370 521 700, clip,scale=0.35]{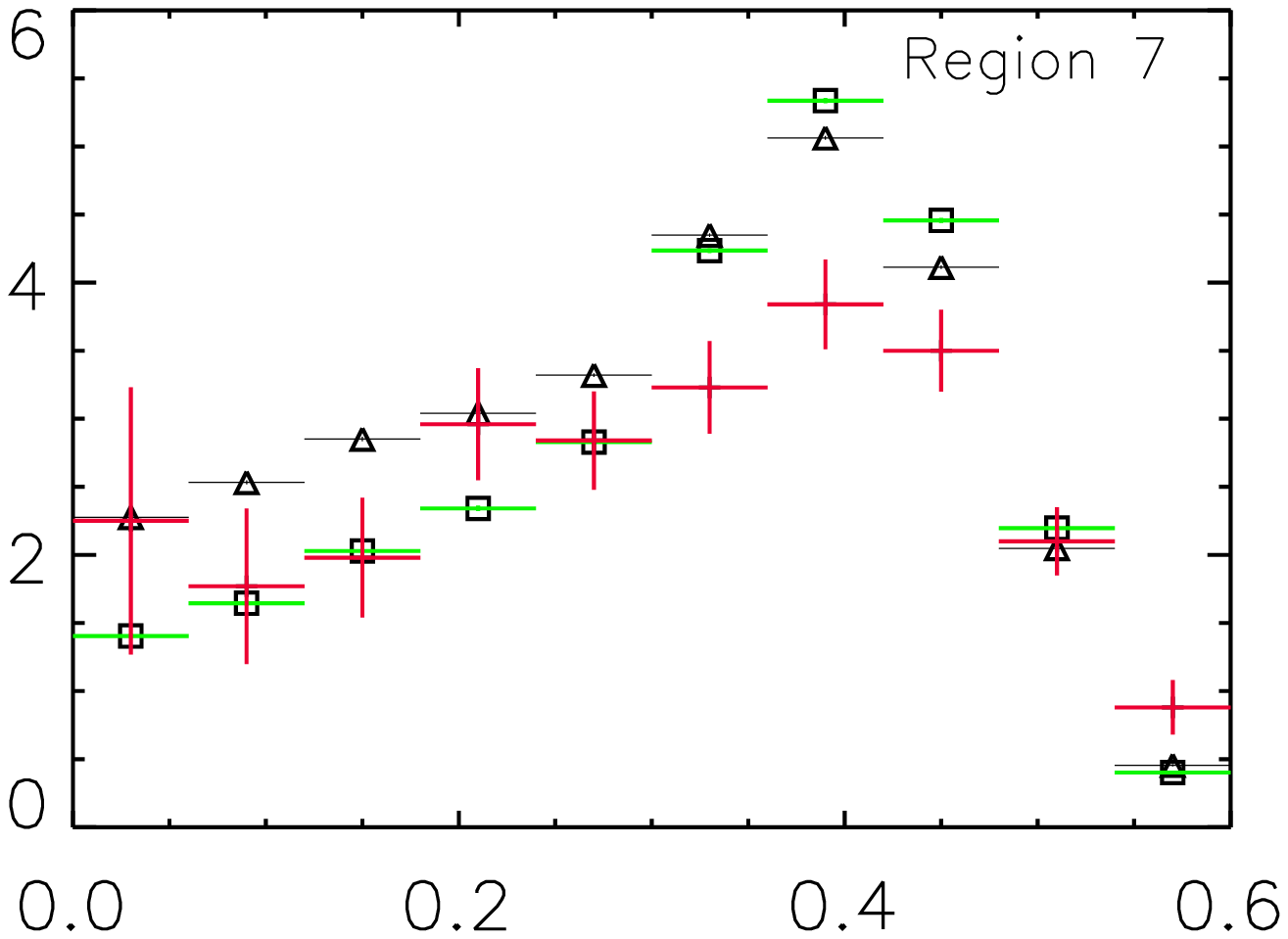} } & \hspace{-7.59mm}
 { \includegraphics[bb= 178  370 521 700, clip,scale=0.35]{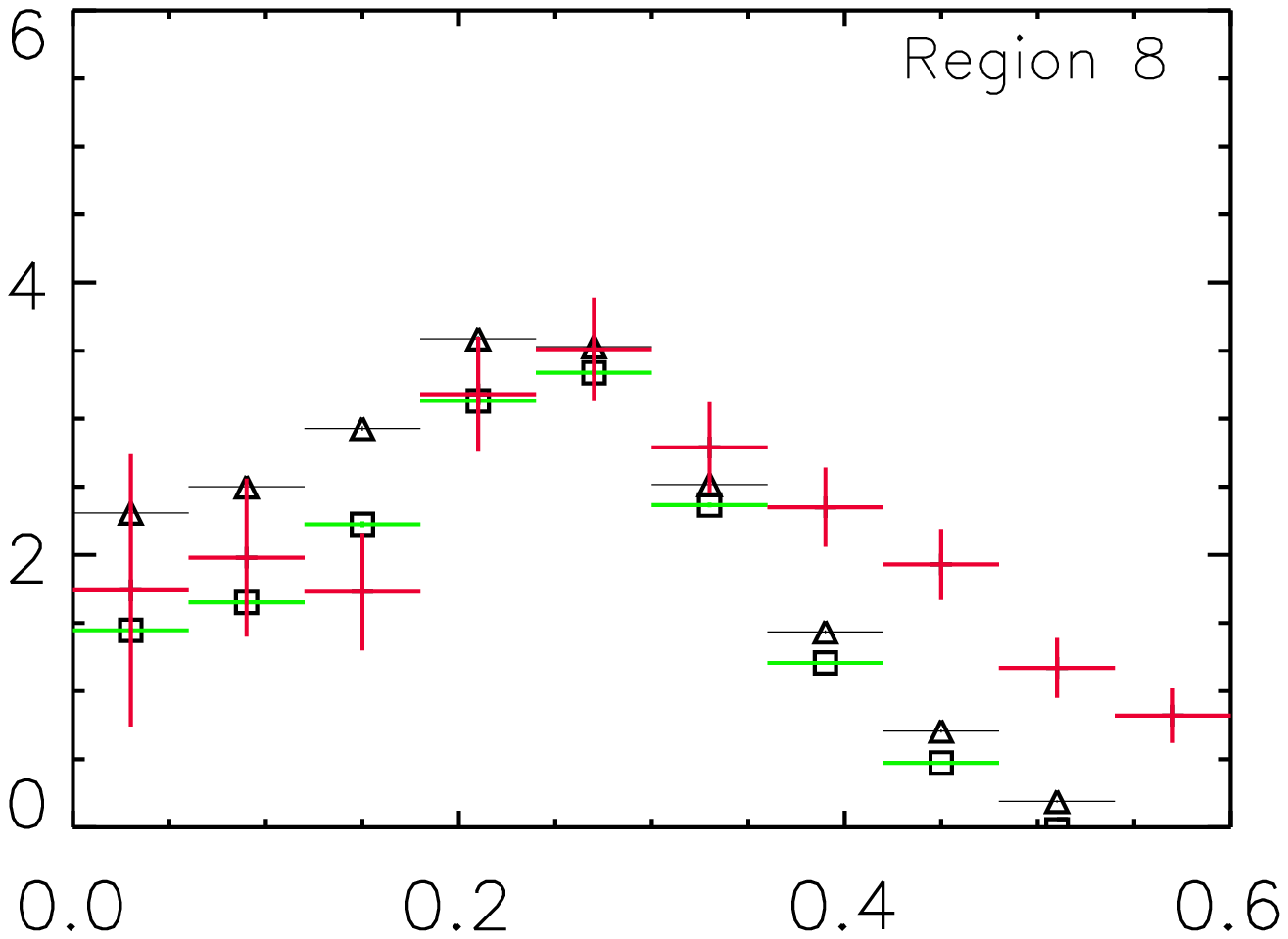} } 

   \end{tabular}

   \caption{ Radial profiles in X-rays in two energy bands and in \g-rays for the 8 sectors defined in Fig. \ref{fig:regioncam}. The general agreement is good
    and particularly striking in sector 7 (the  brightest spot of the remnant in both wavelengths).
 However there are also interesting
    differences in sectors 5  and 6  where the bulk of the X-ray emission seems to come  more from the inside of the SNR than in    \g-rays.}

   \label{fig:radprofiles}
\end{figure*}

The global agreement in the 8 sectors is good and particularly striking in sector 7 (the brightest spot of the remnant
 in both wavelength) as can be seen in Fig. \ref{fig:radprofiles}. In regions 5 and 6 we see that the peak of the emission,
 along  the radial direction, is located closer to the center in X-rays than in \g-rays.
 The global shape of the radial profile is similar in both wavelength but shifted by $\sim$0.1$\degr$ to the center in X-rays.

It is important to note that we did subtract an astrophysical background to the X-ray radial profiles.
To estimate the background, we used the same region as in the mosaic building section (Sect. \ref{sect:mosaic}).

\section{Discussion}
\label{sect:discu}

\subsection{Global synchrotron spectrum}

The radio flux density at 1.4 GHz that we derive in our study (22 Jy $<$ S $<$ 26 Jy)
for the whole remnant is significantly higher than the flux density previously published.
However it is important to note that the flux calculated by \citet{es01} (S=4$\pm$1
Jy) was estimated only  locally for the two bright arcs in the northwestern region. 
Later on,  \citet{ls04} derived 
for these features a flux density of 6.7 $\pm$ 2.0 Jy based on the
improved radio image at 1.4 GHz. 
In X-rays the contribution of the northwestern region accounts
for $\sim$ 20 $\%$ of the total flux in the  0.5-4.5 keV energy band.
If the ratio of synchrotron emission in X-rays to that in radio is more or less constant 
throughout the remnant, the radio flux expected for the whole remnant is of the order 
of  26 Jy\footnote{Without taking into account the thermal emission from Arc 2 that accounts for
about 1.5 Jy out of the 6.7 Jy estimation from \citet{ls04}.} which is compatible with our estimation.
We note that the flux that we derive is about  2 times higher than the flux of 13.4 Jy used in AH06.
This flux density was obtained  assuming that the flux density for the entire remnant was
about twice the flux density value obtained by \citet{ls04} for the northwestern region. 
Our higher new estimate tend to reduce the difference between the measurements and predictions  of
the hadronic models usually  requiring high magnetic fields $>$ 100 $\mu$G \citep{tu08}.\\
In X-rays, the 1-10 keV non absorbed flux that we derive
with \x is significantly lower than the value derived by \citet{tu08} with \textit{Suzaku} 
(5.23 and 7.65  $\times 10^{-10}$  erg cm$^{-2}$s$^{-1}$ respectively, see Sect. \ref{sect:xflux}).
 This new X-ray flux together with the higher radio flux estimate
decrease the global  X-ray to radio ratio by about a factor of 3. Such a change 
is clearly important in the shape of the synchrotron spectral energy distribution 
and could impact the results from multi-wavelength models of the remnant.

\subsection{X- and \g-ray comparison}

\subsubsection{Spectral index}
When using large extraction regions and degrading the spatial resolution
of the X-ray data, we found no significant
spatial variation of the photon index (see Sect. \ref{sect:specresults}).
In other words, the small-scale variations reported by \citet{cdg04} are washed
out at the resolution of the HESS telescope. If there exists small-scale spectral 
variations in \g-rays, one  expect them to be washed out as well with the current HESS resolution. 
However the mean X-ray photon index (2.32)
is slightly steeper than the \g-ray one (2.09).
In the case of IC off the CMB photons for the \g-ray emission, the
mean electron energy required to have a photon at 1 TeV is
$E_{\mathrm{e}}^{\mathrm{IC}}$= 16 TeV
(for IC off IR or optical photons it is of course lower).
The electrons emitting synchrotron  at 1 keV with a magnetic field
of 70 $\mu$G have an energy of $E_{\mathrm{e}}^{\mathrm{sync}}$= 27 TeV.
At those energies the electrons are close to the cutoff and we do expect
a higher index when looking at electrons of higher energies.
We can reasonably set an upper limit to the magnetic field as
$E_{\mathrm{e}}^{\mathrm{sync}}$ cannot be lower than
$E_{\mathrm{e}}^{\mathrm{IC}}$ in order to reproduce the steeper index
of the synchrotron spectra.
This sets an upper limit on the downstream magnetic field of 200 $\mu$G,
consistent with the estimate coming from the width of the X-ray filaments
\citep[about 70 $\mu$G from][]{jb06,bv06}.

\subsubsection{Correlation between the X-ray and \g-ray fluxes.}
\label{sect:Xgcor}

We have found in Sect. \ref{sect:specresults} that the X-ray flux is indeed
correlated with the \g-ray flux, but the correlation appears non-linear,
more like ${\rm d} \log F_\nu^{\rm X} / {\rm d} \log F_\nu^{\rm \gamma }$ =
2.41 $\pm$ 0.55
(Fig. \ref{fig:compspec2}).
We discuss here why it could be so. We will assume that most of
the range in flux that we see is due to density variations around the
remnant. It is certainly not entirely true (geometric effects such
as limb brightening must play a role at some point) but it is probably
indeed the major contributor in view of the very structured interstellar
gas and X-ray image.
We will assume that in the range which emits the \g and X-ray
(\textit{i.e.} one to a few hundred TeV) the
particle distribution can be represented as a cut-off power law
$dN/dE = K E^{-s} C(-E/E_{\rm c})$ in which K is proportional
to the ambient density and $E_{\rm c}$ is limited by synchrotron losses
for electrons ($E_{\rm c,e} \propto V_{\rm sh} / \sqrt{B_{\rm d}}$) 
and by age for protons ($E_{\rm c,p} \propto B_{\rm d} V_{\rm sh}^2$) where 
$B_{\rm d}$ is the downstream magnetic field.
 The shape of the cutoff C(x) can be any function decreasing from 1 to 0.
We will further assume that  magnetic field may increase
with density ($B_{\rm d} \propto n^\beta$) and that the shock velocity
adjusts as $V_{\rm sh} \propto n^{-0.5}$ (same pressure).

By calculating ${\rm d} \log F_\nu / {\rm d} \log n$ for the synchrotron
and Inverse Compton case 
(see Appendix \ref{sect:app1}), we can then predict
in the flux-flux correlation comparable to Fig. \ref{fig:compspec2}  :

\begin{equation}
\frac{{\rm d} \log F_\nu^{\rm sync}}{{\rm d} \log F_\nu^{\rm IC}} =
\frac{\frac{s+1}{2} (1 + \beta) - \alpha_X}
{1 - (1 + \beta) (\alpha_\gamma - \frac{s-1}{2})}
\end{equation}
where $\alpha_X$ and $\alpha_\gamma$ are respectively the X and \g-ray spectral slope in energy not photons
($\alpha_i = - {\rm d} \log F_\nu^{i} / {\rm d} \log \nu$).

For a standard value of $s$ = 2 and the observed values of $\alpha_X$ = 1.32
and $\alpha_\gamma$ = 1.09 this gives
(0.18 + 1.5 $\beta$) / (0.41 - 0.59 $\beta$).
If $B_{\rm d}$ is insensitive to density ($\beta$ = 0) the predicted correlation
is opposite of what is seen: the range in X-ray flux would be smaller
than in \g-rays, because the negative feedback on $\nu_{\rm c}$
via $V_{\rm sh}$ plays more strongly in X-rays which are further
in the cutoff part of the spectrum.
But a modest dependence of $B_{\rm d}$ on density like $\beta$ = 0.1 is enough
to invert the trend because in the loss-dominated regime the \g-ray cutoff
frequency decreases with $B_{\rm d}$ while the X-ray one is independent of $B_{\rm d}$.
In other words, the slope of the log($F_X$) vs log($F_\gamma$)
correlation is very sensitive to $\beta$.
To get the observed value of 2.41 requires $\beta$ = 0.28.

In view of the oversimplified character of that approach we do not claim
that this is a measurement of ${\rm d} \log  B_{\rm d} / {\rm d} \log n$
but we think it shows that
such a steep correlation is reasonable in a leptonic model.
The specific model in which we have tried to push the calculation
further (Appendix \ref{sect:app2}) does not give a consistent answer,
but it is far from unique.
One way to improve on the measurements would be to use extraction areas
in which the filling factor of the SNR is the same, like the angular
sectors on Fig.\ref{fig:regioncam}. 
This would leave in the flux variations only what is due
to varying external conditions.
It requires reanalyzing the HESS
data, so it is left for future work.

If we now turn to the hadronic hypothesis,
the same line of reasoning  (see Appendix \ref{sect:app1}) then leads to :

\begin{equation}
\frac{{\rm d} \log F_\nu^{\rm sync}}{{\rm d} \log F_\nu^{\rm hadr}} =
\frac{\frac{s+1}{2} (1 + \beta) - \alpha_X}
{2 - (1 - \beta) (\alpha_\gamma + 1 - s)}
\label{eq:pi0main}
\end{equation}

For a standard value of $s$ = 2 and the observed values of $\alpha_X$ = 1.32
and $\alpha_\gamma$ = 1.09 this gives
(0.18 + 1.5 $\beta$) / (1.91 + 0.09 $\beta$).
It is clear that whatever $\beta < 1$ (it is hard to imagine how $B_{\rm d}$
could increase faster than density) this quantity is always $<$ 1.
In other words, no magnetic field can make up for the natural $n^2$
character of the hadronic mechanism which predicts a fast increase
of $F_\nu^{\rm hadr}$ with density.
So at least in that (over)simplified framework the correlation we observe
is not in favor of a hadronic model.

There remains the possibility \citep{md05}
that the density increases very fast outwards (SNR hitting a shell)
to the point where most of the \g-ray emission arises outside the remnant
(in the precursor). In that case the width of the precursor increases
as $E$ so that the spectral shape of the \g-ray emission may be estimated
by multiplying  $F_\nu^{\rm hadr}$ by $E$ \citep{za07}.
This in turn changes (\ref{eq:pi0main}) into

\begin{equation}
\frac{{\rm d} \log F_\nu^{\rm sync}}{{\rm d} \log F_\nu^{\rm hadr}} =
\frac{\frac{s+1}{2} (1 + \beta) - \alpha_X}
{2 - (1 - \beta) (\alpha_\gamma + 2 - s)}
\label{eq:pi0main2}
\end{equation}
so that for $s$ = 2 and the observed spectral indices one expects
(0.18 + 1.5 $\beta$) / (0.91 + 1.09 $\beta$)
for the slope of the log($F_X$) vs log($F_\gamma$)
correlation. This is still always $<$ 1.

\subsection{Spatial comparison}

The main difficulty of a leptonic model to account for the observations
in \rxj is that it requires a small magnetic field (on the order of 10 $\mu$G ; AH06)
to explain the rather large \g to X-ray ratio, if the emitting volume is
the same. This is inconsistent with the magnetic field derived from
the width of the X-ray filaments (70 $\mu$G or so).
A possible reason, suggested by \cite{ls04},
is that the magnetic turbulence decays
behind the shock faster than the electrons lose energy. This leaves
a larger volume (downstream) to IC than synchrotron, and does not require
that large a magnetic field to begin with. 
A definite prediction is then that the \g-ray emission should peak inside
the X-rays.

In the comparison of the X- and \g-ray radial profiles
(\ref{sect:radprofresults}) we did see a radial shift,
particularly visible in region 6 (West),
between the X- and \g-ray emission.
But the shift is in the opposite direction, \textit{i.e.}
the X-ray emission peaks at smaller radius than the \g-ray emission.
The value of this shift for region 6
 is $\sim$ 0.06$\degr$=1.2 pc with a remnant at 1 kpc.
Another effect that we have not discussed is that because of the magnetic
jump at the shock (typically a factor 3 if the magnetic field is mostly
turbulent and isotropic), the synchrotron emission is strongly suppressed
ahead of the shock whereas the IC emission will decrease more
smoothly over one diffusion length.
In a Bohm regime the diffusion coefficient is
$D_{\mathrm{B}}=r_{\mathrm{L}}c/3$ where $r_{\mathrm{L}}$
is the Larmor radius and the diffusion length
$l_{\mathrm{diff}}=D_{\mathrm{B}}/V_{\mathrm{sh}}$
where $V_{\mathrm{sh}}$ is the shock wave velocity.
For an electron energy of 16 TeV, a shock speed of 4000 km/s and
an upstream magnetic field of 10 $\mu$G we have
$l_{\mathrm{diff}}=0.1$ pc. This is small in comparison
with the shift of 1.2 pc.

However there exists another purely geometric possibility to explain
why the X-rays peak inside the \g-rays.
Actually from Fig.\ref{fig:radprofiles}
the effect is significant only in regions 6 (West) and 8 (North).
In both regions it looks from the image (Fig.\ref{fig:mosaic}) that the
remnant extends beyond the main X-ray peak. This is typical of a line of sight
superposition of a region with larger density (brighter, slower shock)
and a region of lower density (fainter, faster shock). This
 idea is supported by the fact that observations in CO \citep{mt05} show
 that clouds are present in those particular regions.
Now from Fig.\ref{fig:compspec} we know that the X-ray brightness increases
much faster than the \g-ray brightness. The X-ray radial profile is then
dominated by the inner bright edge, whereas the outer plateau may
contribute significantly to the \g-ray emission and shift the integral
along the line of sight to a larger radius.
Because of that possibility we cannot say that the radial profile
is incompatible with a leptonic model.

In a hadronic model the simplest way to understand that the \g-ray emission
is further out is if a sizable fraction of the flux comes from outside
the shock. This is possible and actually expected \citep{md05} if
the ambient density increases ahead of the shock (SNR entering a molecular
cloud). There are then vast amounts of upstream gas to be used as targets for
accelerated protons. A definite prediction of that model is that the
remnant should look larger as energy increases because it will be
governed by the diffusion length. The energy range in the HESS data
may be enough to test that, and certainly will after HESS 2.

\section{Summary}

New \x observations of \rxj have allowed us to complete the coverage of the remnant.
We have then carried out a comparison of the remnant in X- and \g-rays. The radio emission
has also been studied. This leads us to the following conclusions :

\begin{enumerate}

\item The small scale variations of the X-ray photon index are largely smeared out
at the resolution of the HESS telescope.  This is consistent with the non detection of 
photon index variation in the \g-rays.

\item The mean X-ray photon index (2.32) is slightly steeper than the \g-ray one (2.09).
This is expected in a leptonic scenario as the electrons emitting \g-ray Inverse Compton have a lower
energy than those emitting X-ray synchrotron. The photon index offset is thus the result of
the spectrum cutoff of the electrons.

\item The comparison of the X-ray vs \g-ray integrated flux suggests a non-linear correlation
with $\rm d \log F_\nu^{\rm X} / \rm d \log F_\nu^{\rm \gamma}$ = 2.41 $\pm$ 0.55. If the range of flux that we see
is due to the variation of the density around the remnant then a leptonic model can more easily reproduce the
observed X/\g-ray correlation.

\item The comparison of the radial profile in X- and \g-rays indicates that for some regions,
the X-ray emission comes more from the inside of the remnant than in \g-rays. This can be explained
 in a hadronic model if a fraction of the \g-ray flux comes from the outside of the remnant
 (interaction of the remnant with a cloud). However this radial shift could also be due to
  superposition effects. Therefore we can not rule out the leptonic scenario.

 \item Concerning the radio counterpart of the remnant, we have shown that one of the brightest arcs seen at 1.4 GHz
  (Arc 2) is of thermal origin and likely not associated with the remnant. Taking this into account, 
we have derived a lower limit on the integrated  flux density of 22 Jy and an upper limit of 26 Jy.

\end{enumerate}

\begin{acknowledgements}
This project was partially funded by ECOS-SUD France-Argentina A04U03
program and ANPCYT, CONICET and UBACYT (Argentina) grants.
G. Dubner and E. Giacani are members of CONICET (Argentina). M. Ortega is
a Doctoral Fellow of CONICET (Argentina).
\end{acknowledgements}

\begin{appendix} 
\section{Correlation between the X-ray and \g-ray fluxes}
\label{sect:app1}

The synchrotron emission can be written
\begin{equation}
F_\nu^{\rm sync} \propto K_{\rm e} \, B^{\frac{s+1}{2}} \,
  \nu^{-\frac{s-1}{2}} \: S \left( \frac{\nu}{\nu_{\rm c}^{\rm sync}} \right)
\label{eqn:emsync}
\end{equation}
where $S$ is a function characteristic of synchrotron emission and the shape of the electron cutoff
and $\nu_{\rm c}^{\rm sync} \propto B E_{\rm c,e}^2 
                            \propto V_{\rm sh}^2$.
Then
\begin{equation}
\frac{{\rm d} \log F_\nu^{\rm sync}}{{\rm d} \log n} = 1 + \frac{s+1}{2} \beta
- \frac{S'}{S} \, \frac{\nu}{\nu_{\rm c}^{\rm sync}} \, 
\frac{{\rm d} \log \nu_{\rm c}^{\rm sync}}{{\rm d} \log n}
\end{equation}
Noting that
\begin{equation}
\frac{{\rm d} \log F_\nu^{\rm sync}}{{\rm d} \log \nu} = - \frac{s-1}{2} +
\frac{S'}{S} \, \frac{\nu}{\nu_{\rm c}^{\rm sync}} = - \alpha_X
\end{equation}
is the X-ray spectral slope (in energy, not photons) and that
${\rm d} \log \nu_{\rm c}^{\rm sync} / {\rm d} \log n = -1$, we can write
\begin{equation}
\frac{{\rm d} \log F_\nu^{\rm sync}}{{\rm d} \log n} = 1 + \frac{s+1}{2} \beta
- \alpha_X + \frac{s-1}{2} = \frac{s+1}{2} (1 + \beta) - \alpha_X
\label{eq:syncflux}
\end{equation}

As long as the Klein-Nishina reduction of the cross-section is not reached
(this is still true for 10 TeV electrons on the CMB) the Inverse Compton
emission may be written in the same way as (\ref{eqn:emsync}) but without
any explicit $B$ term and with
$\nu_{\rm c}^{\rm IC} \propto E_{\rm c,e}^2 \propto V_{\rm sh}^2 / B$.
The same line of reasoning then leads to
\begin{equation}
\frac{{\rm d} \log F_\nu^{\rm IC}}{{\rm} d \log n} = 
1 - (1 + \beta) \left( \alpha_\gamma - \frac{s-1}{2} \right)
\label{eq:icflux}
\end{equation}
in which $\alpha_\gamma$ is  the \g-ray spectral slope.

Putting together (\ref{eq:syncflux}) and (\ref{eq:icflux}) we then predict
in the flux-flux correlation comparable to Fig.\ref{fig:compspec2} 
\begin{equation}
\frac{{\rm d} \log F_\nu^{\rm sync}}{{\rm d} \log F_\nu^{\rm IC}} =
\frac{\frac{s+1}{2} (1 + \beta) - \alpha_X}
{1 - (1 + \beta) (\alpha_\gamma - \frac{s-1}{2})}
\end{equation}

If we now turn to the hadronic hypothesis,
the $\pi^0$ decay emission may be written
\begin{equation}
F_\nu^{\rm hadr} \propto K_{\rm p} \, n \, \nu^{1-s} \:
H \left( \frac{\nu}{\nu_{\rm c}^{\rm hadr}} \right)
\label{eqn:empi0}
\end{equation}
where $H$ is a function characteristic of $\pi^0$ emission  and the shape of the proton cutoff  and
$\nu_{\rm c}^{\rm hadr} \propto E_{\rm c,p} \propto B V_{\rm sh}^2$.
The same line of reasoning then leads to
\begin{equation}
\frac{{\rm d} \log F_\nu^{\rm hadr}}{{\rm d} \log n} = 
2 - (1 - \beta) (\alpha_\gamma + 1 - s)
\label{eq:pi0flux}
\end{equation}
in which we now assume that hadronic emission dominates
the \g-ray spectral slope.
Putting together (\ref{eq:syncflux}) and (\ref{eq:pi0flux}) we then predict
for the flux-flux correlation in the hadronic model
\begin{equation}
\frac{{\rm d} \log F_\nu^{\rm sync}}{{\rm d} \log F_\nu^{\rm hadr}} =
\frac{\frac{s+1}{2} (1 + \beta) - \alpha_X}
{2 - (1 - \beta) (\alpha_\gamma + 1 - s)}
\label{eq:pi0corr}
\end{equation}

There remains the possibility \citep{md05}
that the density increases very fast outwards (SNR hitting a shell)
to the point where most of the \g-ray emission arises outside the remnant
(in the precursor). In that case the width of the precursor increases
as $E$ so that the spectral shape of the \g-ray emission may be estimated
by multiplying  \ref{eqn:empi0}  by $E$ \citep{za07}.

This in turn changes (\ref{eq:pi0flux}) into

\begin{equation}
\frac{{\rm d} \log F_\nu^{\rm hadr}}{{\rm d} \log n} = 
2 - (1 - \beta) (\alpha_\gamma + 2 - s)
\label{eq:pi0flux2}
\end{equation}

and changes (\ref{eq:pi0corr}) into

\begin{equation}
\frac{{\rm d} \log F_\nu^{\rm sync}}{{\rm d} \log F_\nu^{\rm hadr}} =
\frac{\frac{s+1}{2} (1 + \beta) - \alpha_X}
{2 - (1 - \beta) (\alpha_\gamma + 2 - s)}
\label{eq:pi0corr2}
\end{equation}

\section{Detailed calculation for electrons dominated by radiative losses}
\label{sect:app2}

If the cutoff in the electron spectrum is defined by synchrotron cooling
as we assume in Sect.\ref{sect:Xgcor} and App.\ref{sect:app1},
then the cooling continues downstream
and the electron distribution integrated over space gets steeper
(the power law index increases by 1) down to a break energy
$E_{\rm b} \propto B^{-2} t_0^{-1}$ in which $t_0$ is the SNR age \citep{za07}.
Since we are interested only in what happens above $E_{\rm b}$ this can be
accounted for in (\ref{eqn:emsync})
by changing $s$ to the steeper value
and considering that the normalization $K_e$ follows $E_{\rm b}$
(the total number of electrons is always dominated by those below $E_{\rm b}$)
or $B^{-2}$ ($t_0$ is the same in all parts of the SNR).
This amounts to adding $- 2 \beta$ to Eqs (\ref{eq:syncflux}) and
(\ref{eq:icflux}), resulting in
\begin{eqnarray}
\frac{{\rm d} \log F_\nu^{\rm sync}}{{\rm d} \log n} =
 \frac{s+1}{2} - \alpha_X + \frac{s-3}{2} \beta \\
\frac{{\rm d} \log F_\nu^{\rm IC}}{{\rm d} \log n} = \frac{s+1}{2} -
 \alpha_\gamma - \beta \left( \alpha_\gamma + \frac{5-s}{2} \right)
\end{eqnarray}

For the most interesting case $s$ = 3 we then have
\begin{equation}
\frac{{\rm d} \log F_\nu^{\rm sync}}{{\rm d} \log F_\nu^{\rm IC}} =
\frac{2 - \alpha_X}
{2 - \alpha_\gamma - \beta (\alpha_\gamma + 1)}
\end{equation}
For $\alpha_X$ = 1.32 and $\alpha_\gamma$ = 1.09 we get
$0.68 / (0.91 - 2.09 \, \beta)$. This is very different from what is obtained
with $s$ = 2 and no break in the electron spectrum, but reaches the observed
value 2.41 for a very similar value of $\beta$ = 0.30.

The pion decay emission is insensitive to synchrotron cooling
so that ``radiative'' model would predict 
${\rm d} \log F_\nu^{\rm sync} / {\rm d} \log F_\nu^{\rm hadr} =
0.68 / (1.91 + 0.09 \, \beta)$.
This is again always less than 1.

The same framework also naturally predicts variations of the spectral index
going with the flux variations. Computing that requires a
specific representation of the spectral shape. Suitable approximate
formulae are given by \citet{za07} in their Eqs (35) and (46).
They apply when there is no jump in magnetic field at the shock.
This is not the most likely situation in our opinion, but it is still
interesting to carry out the exercise to the end.
The observed average spectral slope in X-rays $\alpha_X$ = 1.32
corresponds to $x = \nu / \nu_{\rm c}^{\rm sync} = 4.5$, in keeping with
what was derived by \citet{tu08} ($\nu_{\rm c}^{\rm sync}$ = 0.67 keV).
In the same way, the observed spectral slope in \g-rays $\alpha_\gamma$ = 1.09
corresponds to $y = \nu / \nu_{\rm c}^{\rm IC} = 4.1$.
From Eqs (34) and (45) of \citet{za07} we note that
$\nu_{\rm c}^{\rm IC} / \nu_{\rm c}^{\rm sync}$ = (1.2 TeV) / (2.2 keV)
$(B/100 \mu \rm G)^{-1}$. For $B \simeq 70 \, \mu$G and
$\nu_{\rm c}^{\rm sync}$ = 0.67 keV this predicts
$\nu_{\rm c}^{\rm IC} \simeq$ 0.5 TeV which is reasonable.

Coming back to the spectral index variations,
since $\rm d \log \nu_{\rm c}^{\rm sync} / \rm d \log \textit{n} = -1$
when it is limited by cooling, we have
\begin{eqnarray}
\frac{\rm d \alpha_X}{{\rm d} \log n} = - \frac{\rm d \alpha_X}{{\rm d} \log x}
 \frac{\rm d \log \nu_{\rm c}^{\rm sync}}{{\rm d} \log n} =
 x \frac{\rm d \alpha_X}{{\rm d} x} \\
\frac{\rm d \alpha_X}{{\rm d} \log n} = 0.25 \, x^{0.5} - 0.3795 \, x^{0.6}
 \left(1 + 0.46 \, x^{0.6}\right)^{-2}
\end{eqnarray}
At $x$ = 3.5, this gives ${\rm d} \alpha_X / {\rm d} \log n$ = 0.32.
It then predicts ${\rm d} \log F_\nu^{\rm sync} / {\rm d} \alpha_X$ =
0.74 / 0.32 = 2.3.
The peak to peak dispersion that is observed on the X-ray flux
$\Delta \log F_\nu^{\rm sync} \simeq 1.2$
should then be associated with a peak to peak dispersion on the slope
$\Delta \alpha_X$ = 0.53. This is larger than what is observed
on Fig. \ref{fig:2index} ($\Delta \alpha_X$ = 0.27).
This is qualitatively expected because flux variations are also affected
by the geometry (the regions do not cover the same fraction of the SNR).
However the small amplitude of the index variations implies that
the fraction of the flux variations that is due to geometry
$\Delta \log F_\nu^{\rm geom}$
is large. It should add quadratically with the part due to density
variations ($2.3 \times \Delta \alpha_X$)
so $\Delta \log F_\nu^{\rm geom} \simeq$ 1.17.
Of course geometrical effects should affect the \g-ray  emission
in the same way but the observed dispersion on \g-ray flux
is only $\Delta \log F_\nu^{\rm sync} \simeq 0.6$.
This means we are at a dead-end. That specific model cannot explain
at the same time the relatively uniform X-ray spectra and the larger
contrast in the X-ray flux than the \g-ray flux.

\end{appendix}

\Online

\begin{figure*}
   \centering
\vspace{5cm}
   \includegraphics[trim= 0mm 0mm 0mm 10mm,clip,width=17.5cm]{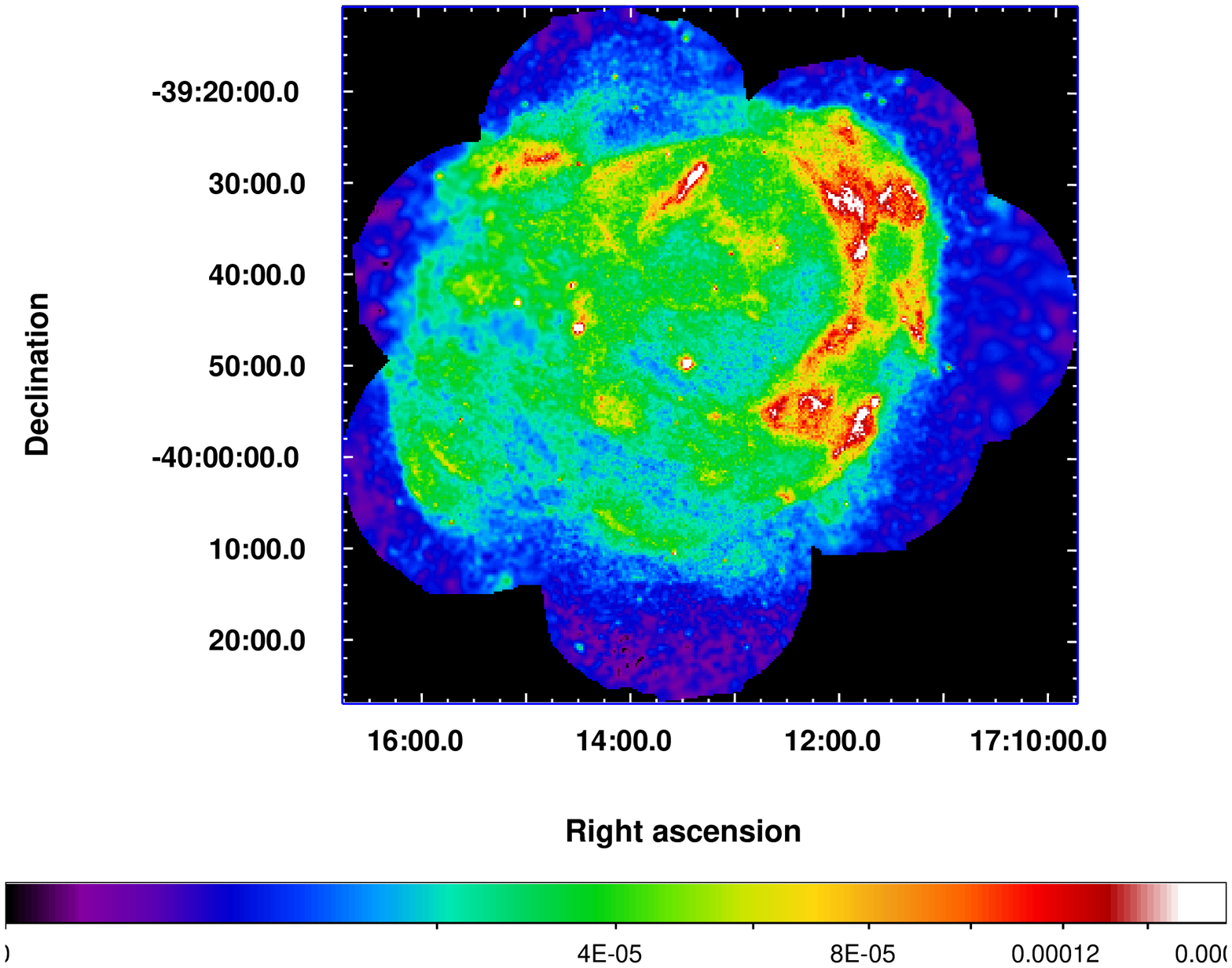}
   \caption{EPIC MOS plus PN image in the 0.5-4.5 keV band. The units are ph/cm$^{2}$/s/arcmin$^{2}$ and 
the scale is square root.  The image was adaptively smoothed to a signal-to-noise ratio of 10. }
   \label{fig:mosaic2}
\end{figure*}


\begin{thebibliography}{31}
\expandafter\ifx\csname natexlab\endcsname\relax\def\natexlab#1{#1}\fi

\bibitem[{{Aharonian} {et~al.}(2006){Aharonian}, {Akhperjanian}, {Bazer-Bachi},
  {Beilicke}, {Benbow}, {Berge}, {Bernl{\"o}hr}, {Boisson}, {Bolz}, {Borrel},
  {Braun}, {Breitling}, {Brown}, {Chadwick}, {Chounet}, {Cornils},
  {Costamante}, {Degrange}, {Dickinson}, {Djannati-Ata{\"i}}, {O'C.~Drury},
  {Dubus}, {Emmanoulopoulos}, {Espigat}, {Feinstein}, {Fontaine}, {Fuchs},
  {Funk}, {Gallant}, {Giebels}, {Glicenstein}, {Goret}, {Hadjichristidis},
  {Hauser}, {Hauser}, {Heinzelmann}, {Henri}, {Hermann}, {Hinton}, {Hofmann},
  {Holleran}, {Horns}, {Jacholkowska}, {de Jager}, {Kh{\'e}lifi}, {Klages},
  {Komin}, {Konopelko}, {Latham}, {Le Gallou}, {Lemi{\`e}re},
  {Lemoine-Goumard}, {Lohse}, {Martin}, {Martineau-Huynh}, {Marcowith},
  {Masterson}, {McComb}, {de Naurois}, {Nedbal}, {Nolan}, {Noutsos}, {Orford},
  {Osborne}, {Ouchrif}, {Panter}, {Pelletier}, {Pita}, {P{\"u}hlhofer},
  {Punch}, {Raubenheimer}, {Raue}, {Rayner}, {Reimer}, {Reimer}, {Ripken},
  {Rob}, {Rolland}, {Rowell}, {Sahakian}, {Saug{\'e}}, {Schlenker},
  {Schlickeiser}, {Schuster}, {Schwanke}, {Siewert}, {Sol}, {Spangler},
  {Steenkamp}, {Stegmann}, {Superina}, {Tavernet}, {Terrier}, {Th{\'e}oret},
  {Tluczykont}, {van Eldik}, {Vasileiadis}, {Venter}, {Vincent}, {V{\"o}lk}, \&
  {Wagner}}]{ah06}
{Aharonian}, F., {Akhperjanian}, A.~G., {Bazer-Bachi}, A.~R., {et~al.} 2006 (AH06),
  \aap, 449, 223

\bibitem[{{Aharonian} {et~al.}(2004){Aharonian}, {Akhperjanian}, {Aye},
  {Bazer-Bachi}, {Beilicke}, {Benbow}, {Berge}, {Berghaus}, {Bernl{\"o}hr},
  {Bolz}, {Boisson}, {Borgmeier}, {Breitling}, {Brown}, {Bussons Gordo},
  {Chadwick}, {Chitnis}, {Chounet}, {Cornils}, {Costamante}, {Degrange},
  {Djannati-Ata{\"i}}, {Drury}, {Ergin}, {Espigat}, {Feinstein}, {Fleury},
  {Fontaine}, {Funk}, {Gallant}, {Giebels}, {Gillessen}, {Goret}, {Guy},
  {Hadjichristidis}, {Hauser}, {Heinzelmann}, {Henri}, {Hermann}, {Hinton},
  {Hofmann}, {Holleran}, {Horns}, {de Jager}, {Jung}, {Kh{\'e}lifi}, {Komin},
  {Konopelko}, {Latham}, {Le Gallou}, {Lemoine}, {Lemi{\`e}re}, {Leroy},
  {Lohse}, {Marcowith}, {Masterson}, {McComb}, {de Naurois}, {Nolan},
  {Noutsos}, {Orford}, {Osborne}, {Ouchrif}, {Panter}, {Pelletier}, {Pita},
  {Pohl}, {P{\"u}hlhofer}, {Punch}, {Raubenheimer}, {Raue}, {Raux}, {Rayner},
  {Redondo}, {Reimer}, {Reimer}, {Ripken}, {Rivoal}, {Rob}, {Rolland},
  {Rowell}, {Sahakian}, {Saug{\'e}}, {Schlenker}, {Schlickeiser}, {Schuster},
  {Schwanke}, {Siewert}, {Sol}, {Steenkamp}, {Stegmann}, {Tavernet},
  {Th{\'e}oret}, {Tluczykont}, {van der Walt}, {Vasileiadis}, {Vincent},
  {Visser}, {V{\"o}lk}, \& {Wagner}}]{ah04}
{Aharonian}, F.~A., {Akhperjanian}, A.~G., {Aye}, K.-M., {et~al.} 2004, \nat,
  432, 75

\bibitem[{Arnaud {et~al.}(2001)Arnaud, Neumann, Aghanim, Gastaud, Majerowicz,
  \& Hughes}]{an01}
Arnaud, M., Neumann, D.~M., Aghanim, N., {et~al.} 2001, A$\&$A, 365, L80

\bibitem[{{Ballet}(2006)}]{jb06}
{Ballet}, J. 2006, Advances in Space Research, 37, 1902

\bibitem[{{Benjamin} {et~al.}(2003){Benjamin}, {Churchwell}, {Babler}, {Bania},
  {Clemens}, {Cohen}, {Dickey}, {Indebetouw}, {Jackson}, {Kobulnicky},
  {Lazarian}, {Marston}, {Mathis}, {Meade}, {Seager}, {Stolovy}, {Watson},
  {Whitney}, {Wolff}, \& {Wolfire}}]{bc03}
{Benjamin}, R.~A., {Churchwell}, E., {Babler}, B.~L., {et~al.} 2003, \pasp,
  115, 953

\bibitem[{{Berezhko} \& {V{\"o}lk}(2006)}]{bv06}
{Berezhko}, E.~G. \& {V{\"o}lk}, H.~J. 2006, \aap, 451, 981

\bibitem[{{Carter} \& {Read}(2007)}]{cr07}
{Carter}, J.~A. \& {Read}, A.~M. 2007, \aap, 464, 1155

\bibitem[{Cassam-Chena{\"{i}} {et~al.}(2004)Cassam-Chena{\"{i}}, Decourchelle,
  Ballet, Sauvageot, Dubner, \& Giacani}]{cdg04}
Cassam-Chena{\"{i}}, G., Decourchelle, A., Ballet, J., {et~al.} 2004, A$\&$A,
  427, 199

\bibitem[{{Conti} \& {Crowther}(2004)}]{cc04}
{Conti}, P.~S. \& {Crowther}, P.~A. 2004, \mnras, 355, 899

\bibitem[{{Ellison} {et~al.}(2001){Ellison}, {Slane}, \& {Gaensler}}]{es01}
{Ellison}, D.~C., {Slane}, P., \& {Gaensler}, B.~M. 2001, \apj, 563, 191

\bibitem[{{Enomoto} {et~al.}(2002){Enomoto}, {Tanimori}, {Naito}, {Yoshida},
  {Yanagita}, {Mori}, {Edwards}, {Asahara}, {Bicknell}, {Gunji}, {Hara},
  {Hara}, {Hayashi}, {Itoh}, {Kabuki}, {Kajino}, {Katagiri}, {Kataoka},
  {Kawachi}, {Kifune}, {Kubo}, {Kushida}, {Maeda}, {Maeshiro}, {Matsubara},
  {Mizumoto}, {Moriya}, {Muraishi}, {Muraki}, {Nakase}, {Nishijima}, {Ohishi},
  {Okumura}, {Patterson}, {Sakurazawa}, {Suzuki}, {Swaby}, {Takano}, {Takano},
  {Tokanai}, {Tsuchiya}, {Tsunoo}, {Uruma}, {Watanabe}, \& {Yoshikoshi}}]{en02}
{Enomoto}, R., {Tanimori}, T., {Naito}, T., {et~al.} 2002, \nat, 416, 823

\bibitem[{{Fukui} {et~al.}(2003){Fukui}, {Moriguchi}, {Tamura}, {Yamamoto},
  {Tawara}, {Mizuno}, {Onishi}, {Mizuno}, {Uchiyama}, {Hiraga}, {Takahashi},
  {Yamashita}, \& {Ikeuchi}}]{fm03}
{Fukui}, Y., {Moriguchi}, Y., {Tamura}, K., {et~al.} 2003, \pasj, 55, L61

\bibitem[{{Hiraga} {et~al.}(2005){Hiraga}, {Uchiyama}, {Takahashi}, \&
  {Aharonian}}]{hu05}
{Hiraga}, J.~S., {Uchiyama}, Y., {Takahashi}, T., \& {Aharonian}, F.~A. 2005,
  \aap, 431, 953

\bibitem[{Koyama {et~al.}(1997)Koyama, Kinugasa, Matsuzaki, Nishiuchi,
  Sugizaki, Torii, Yamauchi, \& Aschenbach}]{kk97}
Koyama, K., Kinugasa, K., Matsuzaki, K., {et~al.} 1997, PASJ, 49, L7

\bibitem[{Lazendic {et~al.}(2003)Lazendic, Slane, Gaensler, Plucinsky, Hughes,
  Galloway, \& Crawford}]{ls03}
Lazendic, J.~S., Slane, P.~O., Gaensler, B.~M., {et~al.} 2003, ApJ, 593, L27

\bibitem[{{Lazendic} {et~al.}(2004){Lazendic}, {Slane}, {Gaensler}, {Reynolds},
  {Plucinsky}, \& {Hughes}}]{ls04}
{Lazendic}, J.~S., {Slane}, P.~O., {Gaensler}, B.~M., {et~al.} 2004, \apj, 602,
  271

\bibitem[{{Malkov} {et~al.}(2005){Malkov}, {Diamond}, \& {Sagdeev}}]{md05}
{Malkov}, M.~A., {Diamond}, P.~H., \& {Sagdeev}, R.~Z. 2005, \apjl, 624, L37

\bibitem[{{McClure-Griffiths} {et~al.}(2005){McClure-Griffiths}, {Dickey},
  {Gaensler}, {Green}, {Haverkorn}, \& {Strasser}}]{mc05}
{McClure-Griffiths}, N.~M., {Dickey}, J.~M., {Gaensler}, B.~M., {et~al.} 2005,
  \apjs, 158, 178

\bibitem[{{Moriguchi} {et~al.}(2005){Moriguchi}, {Tamura}, {Tawara}, {Sasago},
  {Yamaoka}, {Onishi}, \& {Fukui}}]{mt05}
{Moriguchi}, Y., {Tamura}, K., {Tawara}, Y., {et~al.} 2005, \apj, 631, 947

\bibitem[{{Muraishi} {et~al.}(2000){Muraishi}, {Tanimori}, {Yanagita},
  {Yoshida}, {Moriya}, {Kifune}, {Dazeley}, {Edwards}, {Gunji}, {Hara}, {Hara},
  {Kawachi}, {Kubo}, {Matsubara}, {Mizumoto}, {Mori}, {Muraki}, {Naito},
  {Nishijima}, {Patterson}, {Rowell}, {Sako}, {Sakurazawa}, {Susukita},
  {Tamura}, \& {Yoshikoshi}}]{mu00}
{Muraishi}, H., {Tanimori}, T., {Yanagita}, S., {et~al.} 2000, \aap, 354, L57

\bibitem[{{Pannuti} {et~al.}(2003){Pannuti}, {Allen}, {Houck}, \&
  {Sturner}}]{pa03}
{Pannuti}, T.~G., {Allen}, G.~E., {Houck}, J.~C., \& {Sturner}, S.~J. 2003,
  \apj, 593, 377

\bibitem[{{Pfeffermann} \& {Aschenbach}(1996)}]{pa96}
{Pfeffermann}, E. \& {Aschenbach}, B. 1996, in Roentgenstrahlung from the
  Universe, ed. H.~U. {Zimmermann}, J.~{Tr{\"u}mper}, \& H.~{Yorke}, 267--268

\bibitem[{{Pratt} \& {Arnaud}(2002)}]{pa02}
{Pratt}, G.~W. \& {Arnaud}, M. 2002, \aap, 394, 375

\bibitem[{{Reach} {et~al.}(2006){Reach}, {Rho}, {Tappe}, {Pannuti}, {Brogan},
  {Churchwell}, {Meade}, {Babler}, {Indebetouw}, \& {Whitney}}]{rr06}
{Reach}, W.~T., {Rho}, J., {Tappe}, A., {et~al.} 2006, \aj, 131, 1479

\bibitem[{Rothenflug {et~al.}(2004)Rothenflug, Ballet, Dubner, Giacani,
  Decourchelle, \& Ferrando}]{rb04}
Rothenflug, R., Ballet, J., Dubner, G., {et~al.} 2004, A$\&$A, 425, 121

\bibitem[{{Russeil}(2003)}]{ru03}
{Russeil}, D. 2003, \aap, 397, 133

\bibitem[{{Slane} {et~al.}(1999){Slane}, {Gaensler}, {Dame}, {Hughes},
  {Plucinsky}, \& {Green}}]{sl99}
{Slane}, P., {Gaensler}, B.~M., {Dame}, T.~M., {et~al.} 1999, \apj, 525, 357

\bibitem[{{Tanaka} {et~al.}(2008){Tanaka}, {Uchiyama}, {Aharonian},
  {Takahashi}, {Bamba}, {Hiraga}, {Kataoka}, {Kishishita}, {Kokubun}, {Mori},
  {Nakazawa}, {Petre}, {Tajima}, \& {Watanabe}}]{tu08}
{Tanaka}, T., {Uchiyama}, Y., {Aharonian}, F.~A., {et~al.} 2008, \apj, 685, 988

\bibitem[{{Testori} {et~al.}(2001){Testori}, {Reich}, {Bava}, {Colomb},
  {Hurrel}, {Larrarte}, {Reich}, \& {Sanz}}]{tr01}
{Testori}, J.~C., {Reich}, P., {Bava}, J.~A., {et~al.} 2001, \aap, 368, 1123

\bibitem[{Wang {et~al.}(1997)Wang, Qu, \& Chen}]{wq97}
Wang, Z.~R., Qu, Q.~Y., \& Chen, Y. 1997, A$\&$A, 318, L59

\bibitem[{{Zirakashvili} \& {Aharonian}(2007)}]{za07}
{Zirakashvili}, V.~N. \& {Aharonian}, F. 2007, \aap, 465, 695

\end{thebibliography}
\end{document}